\documentclass[preprint,longbibliography]{revtex4-1}

\usepackage{cancel}

\usepackage{xcolor}
\usepackage{amsmath,amssymb,graphicx}
\usepackage{graphicx}
\usepackage{epstopdf}
\usepackage{mathtools}
\usepackage{upgreek}
\usepackage{notes2bib}
\usepackage{bm}
\usepackage{bm}
\usepackage{color}
\usepackage{braket}
\usepackage{textgreek}
\usepackage{soul}
\usepackage{lipsum}
\usepackage{lineno}
\setstcolor{red}
\usepackage[normalem]{ulem} 
\usepackage{filecontents}

\begin{document}

\title{Linear response theory of open systems with exceptional points}



\author{A. Hashemi$^{1,2,*}$}
\author{K. Busch$^{3,4}$}
\author{D. N. Christodoulides$^5$}
\author{S. K. Ozdemir$^6$}
\author{and R. El-Ganainy$^{1,2,\dagger}$}

\affiliation{$^1$ Department of Physics, Michigan Technological University, Houghton, Michigan, 49931, USA}
\affiliation{$^2$ Henes Center for Quantum Phenomena, Michigan Technological University, Houghton, Michigan, 49931, USA}
\affiliation{$^3$ Humboldt-Universit\"{a}t zu Berlin, Institut f\"{u}r Physik, AG Theoretische Optik \& Photonik, D-12489 Berlin, Germany}
\affiliation{$^4$ Max-Born-Institut, Max-Born-Stra{\ss}e 2A, 12489 Berlin, Germany}
\affiliation{$^5$ CREOL/College of Optics and Photonics, University of Central Florida, Orlando, Florida 32816, USA}
\affiliation{$^6$ Department of Engineering Science and Mechanics, and Materials Research Institute, The Pennsylvania State University, University Park, Pennsylvania 16802, USA}
\affiliation{$^*$ \textrm{Corresponding author: hashemis@mtu.edu}}
\affiliation{$^{\dagger}$ \textrm{Corresponding author: ganainy@mtu.edu}}

\begin{abstract}
Understanding the linear response of any system is the first step towards analyzing its linear and nonlinear dynamics, stability properties, as well as its behavior in the presence of noise. In non-Hermitian Hamiltonian systems, calculating the linear response is complicated due to the non-orthogonality of their eigenmodes, and the presence of exceptional points (EPs). Here, we derive a closer form series expansion of the resolvent associated with an arbitrary non-Hermitian system in terms of the ordinary and generalized eigenfunctions of the underlying Hamiltonian. This in turn reveals an interesting and previously overlocked feature of non-Hermitian systems, namely that their lineshape scaling is dictated by how the input (excitation) and output (collection) profiles are chosen. In particular, we demonstrate that a configuration with an EP of order $M$ can exhibit a Lorentzian response or a super-Lorentzian response of order $M_s$ with $M_s=2,3,\ldots,M$, depending on the choice of input and output channels.


\end{abstract}


\maketitle

\noindent \textbf{INTRODUCTION}\\
Resonance is a universal physical phenomenon that takes place in a large variety of systems across a wide range of spatial and time scales. In optics, the rapid progress in modeling and fabrication has enabled the realization of several photonic resonator structures that can trap light for very long times (high quality factors) \cite{Vahala2003Nature} or confine it in smaller domains compared with the free-space wavelength (nano-scale mode volumes) \cite{Dimitri2010NaturePhotonics}. These devices have become indispensable components in almost every field of optical science and engineering including but not limited to lasers, nonlinear optics, optical communication, quantum optics, and biophotonics. While the notion of a completely closed resonator is sometimes used as an idealization to simplify the analysis and to gain intuition, it is neither realistic nor desirable. Quite the opposite, it is necessary to have open channels between the interior of the resonator and its surrounding in order to facilitate the input/output coupling of light. Thus, even in the absence of material loss, optical resonators are fundamentally non-Hermitian--- a fact that is often overlooked despite some early works that considered non-Hermitian effects in optical systems \cite{Petermann1979IJQE,Siegman1989PRA}. These studies demonstrated how non-Hermitian effects in lasers leave their fingerprint on emission linewidth. Those early works, however, focused on situations where the system does not exhibit spectral singularities known as exceptional points (EPs) since this scenario was not relevant to the experimental setups under study at that time. \\

In recent years, the interest in non-Hermitian optical structures has acquired a new dimension following a number of theoretical studies of parity-time (PT) symmetry in optics \cite{ElGanainy2007P, Musslimani2008O, ElGanainy2008B} and its first experimental demonstrations \cite{Gua2009PRL,ElGanainy2008OB}. This in turn has initiated intense theoretical \cite{Schomerus2010PhysRevLettQNS, Wiersig2011PhysRevANPC, Schomerus2013OptLettTPM, Schomerus2014PhysRevANHT, Jing2014SPL, Wiersig2014ESF, Zhong_2016NJP, Wiersig2016SOEP, Lin2016PhysRevLettESE, Lu2017PhysRevApplied.8.044020, Pick2017, Zhong2018naturecomm, Jiang2018PhysRevApplied.10.064037, Zhong2019SES, Arkhipov2019PhysRevA.99.053806, Bliokh2019NatComm, Franke2019PhysRevLettQQM, Rivero2019PhysRevATRI, Resendiz2020PhysRevResearchTPN, Qi2020PRApplied, Zhong2020PhysRevLett.125.203602, Amin2020PRR, Tzortzakakis2020PhysRevBNHD, Komis2020PhysRevEEIW, Tzortzakakis2020PhysRevASPB, Kristensen2020AdvOptPhotonMER, Khanbekyan2020PhysRevResearchDSS, Hadad2020PFI, Hashemi2021APL, Rivero2021PhysRevLettNHG}, and experimental \cite{Feng2013NatureMaterials, Peng2014NaturePhysics,Hodaei2014PTS, Feng2014SML, Chang2014NatPhotonicsPTS, Peng2014ScienceLIS, Brandstetter2014NatCommRPD, Gao2015NatureONH, Zhen2015NatureSRE, Peng2016CM, Xu2016TET, Doppler2016NatureDEE, Chen2017EPHS, Yoon2018NatureTAL, Zhang2018PLO, Zhao2018THS, Lai2019NatureOEP, Chen2020NatPhysRMD, Xia2021ScienceNTP} investigations of non-Hermitian effects in photonic platforms. In contrast to earlier studies, the notion of EPs is at the heart of these recent works. This in turn has initiated feverish efforts seeking to explore the exotic features of wave dynamics in waveguide and resonator geometries that exhibit EPs. For recent reviews, see \cite{ElGanainy2018PRL,Feng2017JMP,Ozdemir2019JPL,Miri2019JPA,Ramy2019CommPhys, Wiersig2020PhotonResREP}. Despite these intense activities on the one hand, and the fact that the mathematics of non-self adjoint operators is well developed on the other, some of the basic features of complex photonic structures that are pertinent to their non-Hermitian nature are either underestimated or misunderstood. Particularly, on one side of the spectrum, the prevailing traditional point of view treats openness only as a source of energy loss or gain, and hence relies completely on Hermitian intuition to analyze the system. On the other side, some works that deal with non-Hermitian systems exhibiting EPs tend to assume that the linear response associated with a defective Hamiltonian (i.e. a Hamiltonian whose spectrum  has an EP or more) can be studied only within the context of perturbation theory which is conceptually misleading even when the final results are mathematically correct. 

\begin{figure}
	\includegraphics[width=5in]{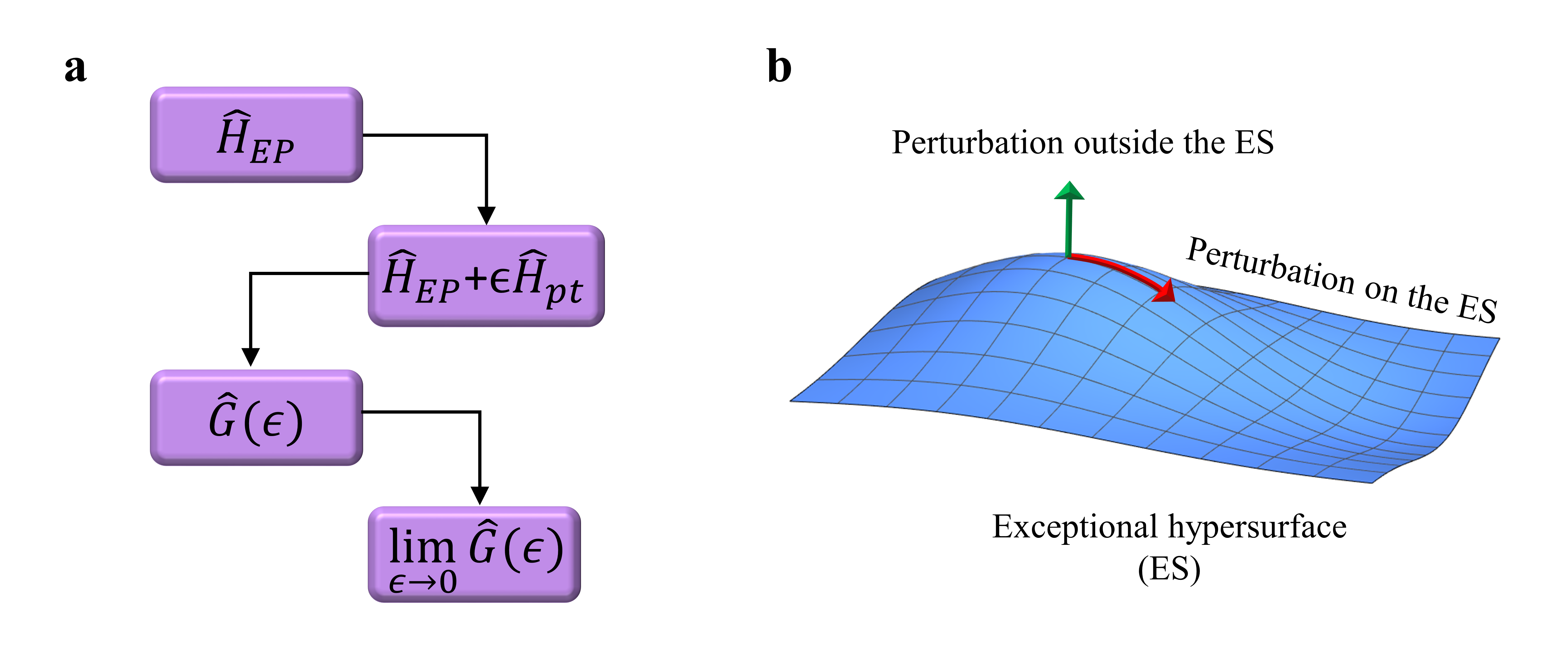}
	\caption{\textbf{Perturbation analysis of defective Hamiltonians ($\hat{H}_{EP}$) and its subtleties at exceptional surfaces.} Two crucial steps are involved: (1) finding a perturbation ($\hat{H}_{pt}$) that removes the degeneracy, and (2) obtaining the limit when the perturbation vanishes. $\hat{G}(\epsilon)$ is the Green operator. The first step can be straightforward in simple systems. \textbf{b} In systems that exhibit exceptional hypersurfaces in the parameter space \cite{Zhong2019SES, ZhouOptica2019, ZhangPhysRevLett2019, soleymani2021chiral, QinQingLaserPhotonicsRev2021}, finding such a perturbation can be a very complex task since any perturbation that shifts the system along the surface will fail. In addition, taking the limit when $\epsilon \rightarrow0$ involves the cancellation of several singular terms with opposite signs. For complex geometries with a large number of degrees of freedom, this can be a daunting task. In general, it is not possible a priori to confirm if this approach gives exact answer or approximate response function.}
	\label{Fig:EP_Surface}
\end{figure}

Beyond this fundamental issue, the perturbative analysis can be cumbersome and complex for non-Hermitian arrangements with large number of resonant elements, particularity when the spectrum contains several EPs, some of which exhibit higher order. This situation becomes relevant for example when studying non-Hermitian topological arrangements \cite{Gong2018PhysRevX.8.031079, Ge2019PhysRevB.100.054105, Liu2019PhysRevLett.122.076801, Liu2020PhysRevB.102.235151}, non-Hermitian disordered media \cite{Tang2020PhysRevA.101.063612, Makris2020PhysRevB.101.014202, Makris2021PhysRevResearch.3.013208}, as well as non-Hermitian spin systems where the number of EPs scale exponentially with the system's size \cite{Luitz2019PhysRevResearch.1.033051}. In addition, the recent discovery of non-Hermitian systems that exhibit exceptional surfaces (ESs) rather than EPs introduces another hurdle for applying perturbation expansions. Figure \ref{Fig:EP_Surface}\textbf{b} illustrates this point. The perturbation analysis used to study a defective Hamiltonian $\hat{H}_{EP}$ typically starts by introducing a perturbation Hamiltonian $\epsilon \hat{{H}}_{pt}$ that removes the non-Hermitian degeneracy. The resolvent of the resultant non-defective Hamiltonian $\hat{\mathcal{H}}_{tot}\equiv \hat{H}_{EP}+\epsilon \hat{{H}}_{pt}$, defined as $G(\omega;\epsilon)\equiv\left(\omega \hat{I}-\hat{\mathcal{H}}_{tot} \right)^{-1}$ where $\omega$ is the frequency and $\hat{I}$ is the unit operator/matrix, can be then obtained using the left and right eigenstates.  The resolvent of the defective Hamiltonian is obtained by evaluating $G_{EP}(\omega)= \lim_{\epsilon\rightarrow 0} G(\omega;\epsilon)$ . Thus, in systems exhibiting exceptional surfaces, one first has to identify the hypersurface of all EPs and carefully identify perturbation Hamiltonians that forces the system out of this hypersurface, otherwise the perturbation analysis will fail since $\hat{\mathcal{H}}_{tot}$ will be also defective (see Fig. \ref{Fig:EP_Surface}\textbf{b}). This task is highly non-trivial since, for systems with large degrees of freedom, the exceptional surface is embedded in a space of high dimensionality. Importantly, taking the above limit involves cancellations of several infinite terms. As a result, any finitely small approximation in applying the perturbation analysis can lead to inconsistent results.\\

In addition to the above mathematical difficulties in using perturbative expansions to analyze defective Hamiltonians, the outcome of this analysis does not provide much insight into the role of non-Hermiticity in shaping the linear response of the system, specially when the latter exhibits many degrees of freedom and several input and output channels. In order to illustrate some of the subtleties arising in such systems in an intuitive way, and in doing so motivates our work, we consider the example shown in Fig. \ref{Fig_Three_Microring}. It consists of three microring resonators that are coupled sequentially via horizontal waveguides. An additional vertical waveguide provides access to selectively excite the second resonator. We neglect the cross talk between the horizontal and vertical waveguides since it can be minimized using various design strategies \cite{Daly1996JLT,Mingaleev2004OptLett,Kobayashi2005OR,Longhi2015OptLett}. A similar system was considered in \cite{Qi2020PRApplied} and shown to exhibit a third-order EP in the subspace spanned by the CW, CCW and CW modes of the resonators $\text{R}_{1,2,3}$, respectively. When the input/output channels are selected as shown in Fig. \ref{Fig_Three_Microring}\textbf{a}, light will cross only cavity $R_3$ and hence the response features a Lorentzian function. On the other hand, for the channels depicted in Fig. \ref{Fig_Three_Microring}\textbf{b}, input light will interact with both cavities $R_{2,3}$ in a series fashion before it exists. One thus expects a super-Lorentzian response of order two. Finally, for the input/output choice shown in Fig. \ref{Fig_Three_Microring}\textbf{c}, light will traverse all three cavities in series which results in a super-Lorentzian response of order three (see supplementary note 1 for detailed analysis of this example). This rather intuitive example reveals that a system with EP of order three can exhibit very different response lineshapes depending on the choice of the input/output channel configuration- a feature that to the best of our knowledge has not been identified in non-Hermitian systems. Needless to say that in more complex structures, identifying the response corresponding to a given input/output channel configuration is not a trivial task. 

\begin{figure}
	\includegraphics[width=\textwidth]{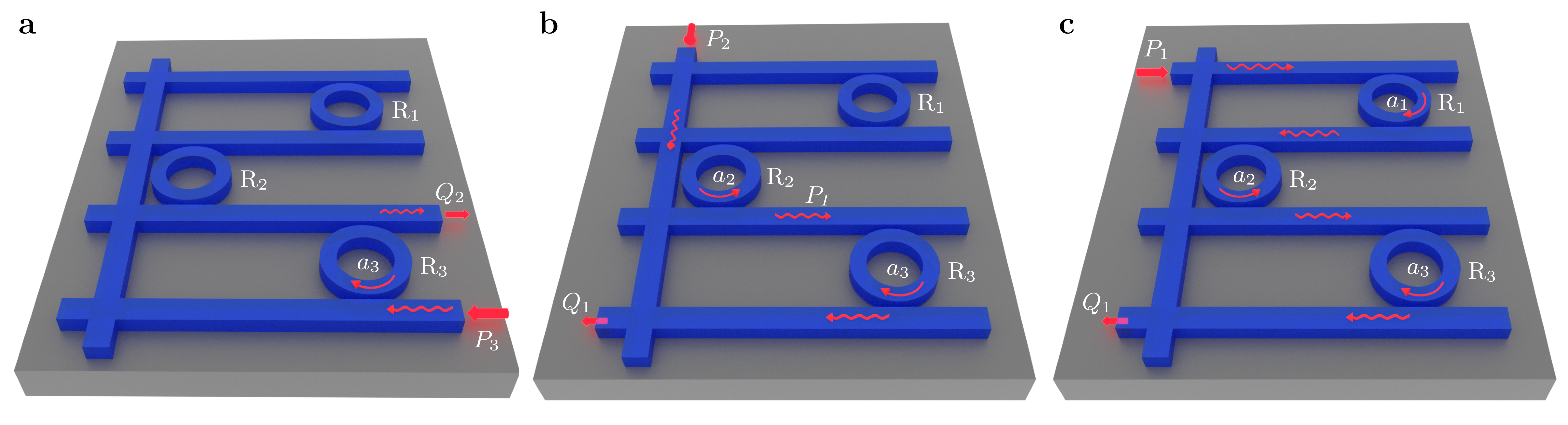}
	\caption{\textbf{Subtleties of linear response of non-Hermitian systems having EPs: an illustrative example.} \textbf{a} An excitation ($P_3$) and collection ($Q_3$) scheme that leads to linear responses featuring Lorentzian lineshape. \textbf{b} and \textbf{c} Excitation ($P_{2,1}$) and collection ($Q_{2,1}$) schemes featuring super-Lorentzian responses of order two and three, respectively. Intuitively the order of the super-Lorentzian response depends on how many cavities the wave has traversed between the input and output ports. $a_{1,2,3}$ are cavity modes for ring resonators $R_{1,2,3}$, respectively.}
	\label{Fig_Three_Microring}
\end{figure}

The lesson learned from the above simple example is of extreme importance since many of the exotic and useful features of non-Hermitian systems with EPs arise due to the modified spectral lineshape. For instance, the recent work on EP-based optical amplifiers shows that the gain-bandwidth product of a resonant optical amplifier can be enhanced by operating at an EP, provided that the response lineshape features a super-Lorentzian response, with better results obtained for higher-order super-Lorentzians \cite{Qi2020PRApplied}. In other settings, super-Lorentzian response could lead to narrower lineshapes, and eventually resulting in to a stronger light-matter coupling, which can be utilized to enhance spontaneous emission \cite{Pick2017, Amin2020PRR,  ren2021quasinormal} and energy harvesting \cite{Fernandez2021} among other potential applications. At the fundamental level, probing the quantum noise in non-Hermitian systems requires a proper characterization of the lineshape response at various output ports due to vacuum fluctuations-induced noise at the input channels, including loss and gain ones \cite{Schomerus_PRA2011}.Thus, in light of the above observation, it will be useful to develop a systematic, generic approach that establishes a universal relation between the response function and the input/output channel configuration.\\

\begin{figure}
\includegraphics[width=5in]{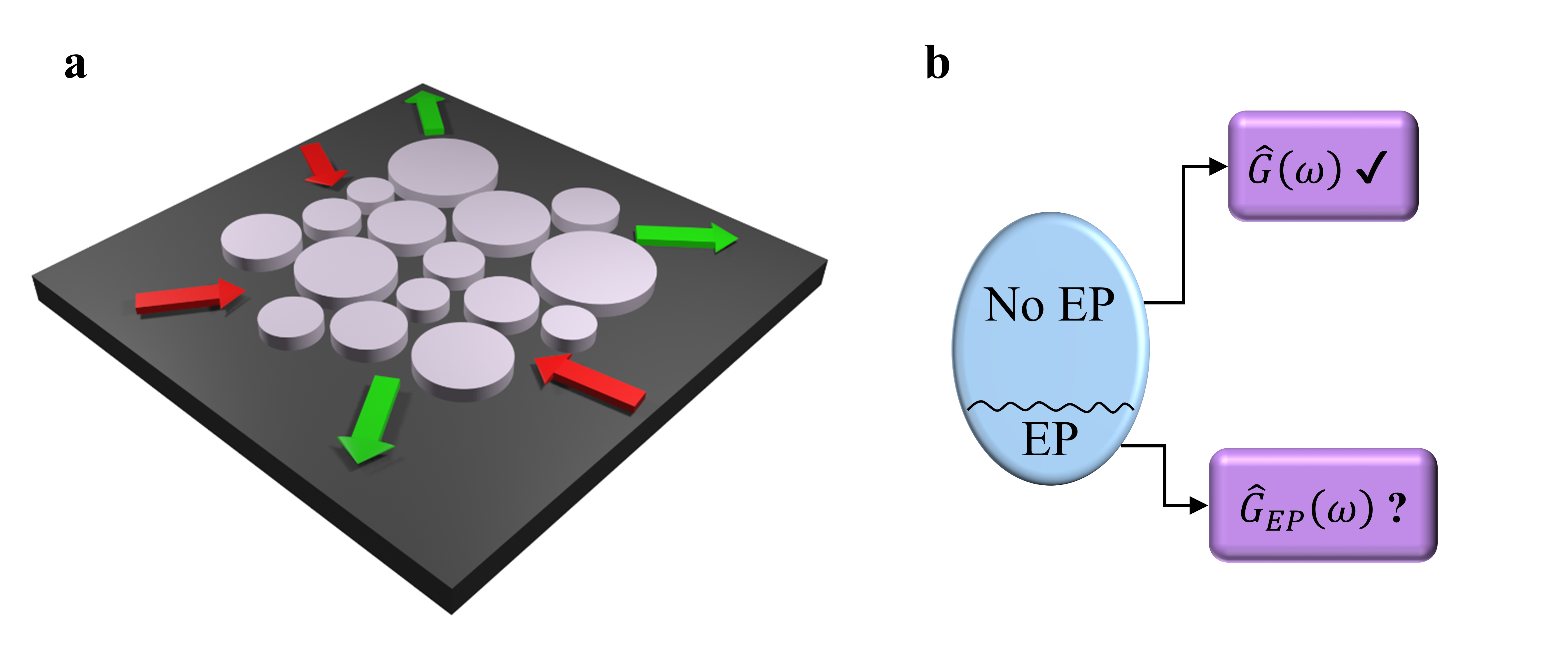}
\caption{\textbf{Non-Hermitian resonant systems and their responses.} \textbf{a} A schematic of a resonant open system, where non-Hermiticity arises due to coupling to external channels and the presence of optical loss or gain. As summarized in \textbf{b}, the linear response of this system, as described by its resolvent operator $\hat{G}(\omega)$ can be readily obtained in terms of the right/left eigenvectors of the associated non-Hermitian Hamiltonian when the spectrum of the latter does not contain any EPs. On the contrary, to date, a general expression for $\hat{G}_{EP}(\omega)$ in the presence of EPs has not been yet developed.}
\label{Fig:Complex_System}
\end{figure}

In this work, we bridge this gap and present a clear and general analysis of the linear response of any non-Hermitian resonant system. Figure \ref{Fig:Complex_System}\textbf{a} presents a schematic of one such a generic system, which consists of large number of coupled non-Hermitian resonators subject to an arbitrary specific choice of input/output channels. It is important to reiterate here that we focus on situations where the eigenvalue spectrum of the system exhibits EPs. While this situation represents a subset of the more general non-Hermitian family of Hamiltonians (see Fig. \ref{Fig:Complex_System}\textbf{b}), it is by now understood that most of the novel behavior of non-Hermitian systems arise when the system is at or near these EP singularities. In addition, in the absence of EPs, the system's response can be easily obtained in terms of the left and right eigenvectors of the Hamiltonian. It is in the case when EPs present that the dimensionality of the eigenspace collapses which complicates the analysis, thus prompting several authors to use perturbation methods as we described earlier. The main results of this work can be summarized as follows: (1) The linear response of resonator geometries that exhibit EPs can be obtained exactly without the need for perturbative expansions; (2) The Green's function expansion can be used to tailor the response of the system by carefully selecting the input/output channels; (3) The excitation channels can be classified based on their interference properties; (4) The most efficient drive of the signal does not necessarily correspond to mode matching between the input signal and resonant modes. Importantly, we emphasize that even though we focus here on optical setups, due to the well-established mathematical analogy between this latter and other physical systems our results will be useful in analyzing and understanding other non-Hermitian platforms such as electronic \cite{2011PhysRevA.84.040101, Schindler_2012, ChenNatElec2018, XiaoPhysRevLett.123.213901, 2019PhysRevLett.123.193901}, acoustic \cite{Pannatoni2011, FleuryNatComm2016, PhysRevB.98.161109, MaNatRevPhys2019, ChenPhysRevApplied2020}, mechanical \cite{SusstrunkE4767, Shomerus_2020_PRR, gupta2022reconfigurable}, and thermal \cite{Fernandez2021} systems.\\

\noindent \textbf{RESULTS} \\

\textit{\textbf{High-$Q$ resonators can be strongly non-Hermitian }}---
Before we proceed to the main topic of this work, it is instructive to first emphasize an important point that is sometimes overlooked in the literature, namely that non-Hermitian effects can be significant even in optical resonators with high quality ($Q$) factors (see for instance \cite{Peng2014NaturePhysics, Wiersig2008PhysRevA}). In general, the degree of non-Hermiticity can be assessed only by comparing the spatial scale of any non-Hermitian perturbation and its strength to the wavelengths of the resonant modes and their spectral separation, respectively. More specifically, if the non-Hermitian perturbation is spatially inhomogeneous and varies at a length scale comparable or smaller than the resonant wavelength, it can result in a measurable discrepancy between the lifetimes associated with different modes. In addition, the perturbation can also introduce hybridization between the bare modes. If such a modal mixing is induced by a coupling strength that is comparable to the frequency difference between the otherwise unperturbed modes, it could bring the system close to an EP. In that case, two or more of the eigenvectors of the system become very close (almost parallel). Obviously, in such a case, non-Hermitian effects beyond linewidth broadening cannot be neglected. A simple example that illustrates this point is the archetypal dimer configuration composed of two resonators having identical resonant frequency $\omega_o$ and two different loss factors $\gamma_{1,2}$. This system is described by a $2 \times 2$ Hamiltonian $H$ with $H_{11,22}=\omega_o-i\gamma_{1,2}$ and $H_{12}=H_{21}=\kappa$ (we assume real coupling). If $\kappa=\Delta \gamma/2$ with $\Delta \gamma=(\gamma_2-\gamma_1)$, the dimer will exhibit an EP. The quality factor associated with this degenerate mode is $Q=\frac{\omega_o}{2\pi \bar{\gamma}},$ where $\bar{\gamma}=(\gamma_1+\gamma_2)/2$. The above formulas do not impose any restrictions on $Q$, which can be engineered to be very large. Yet the system is highly non-Hermitian. The key observation here is that the perturbation $\Delta \gamma$, which introduces additional loss, is spatially inhomogeneous (affects only one site) and is comparable to the energy splitting of the unperturbed system $2\kappa$. Examining the limit $\Delta \gamma=0$ reveals that the eigenstates are orthogonal and the system does not exhibit any drastic non-Hermitian effects even if $Q$ is designed to have a very low value. Thus, the important message here is that the degree of openness alone is insufficient to quantify if the system is highly non-Hermitian or not. This in turn highlights the need to exercise extreme caution when dealing with non-Hermitian systems. \\

\begin{figure}
	\includegraphics[width=3.5in]{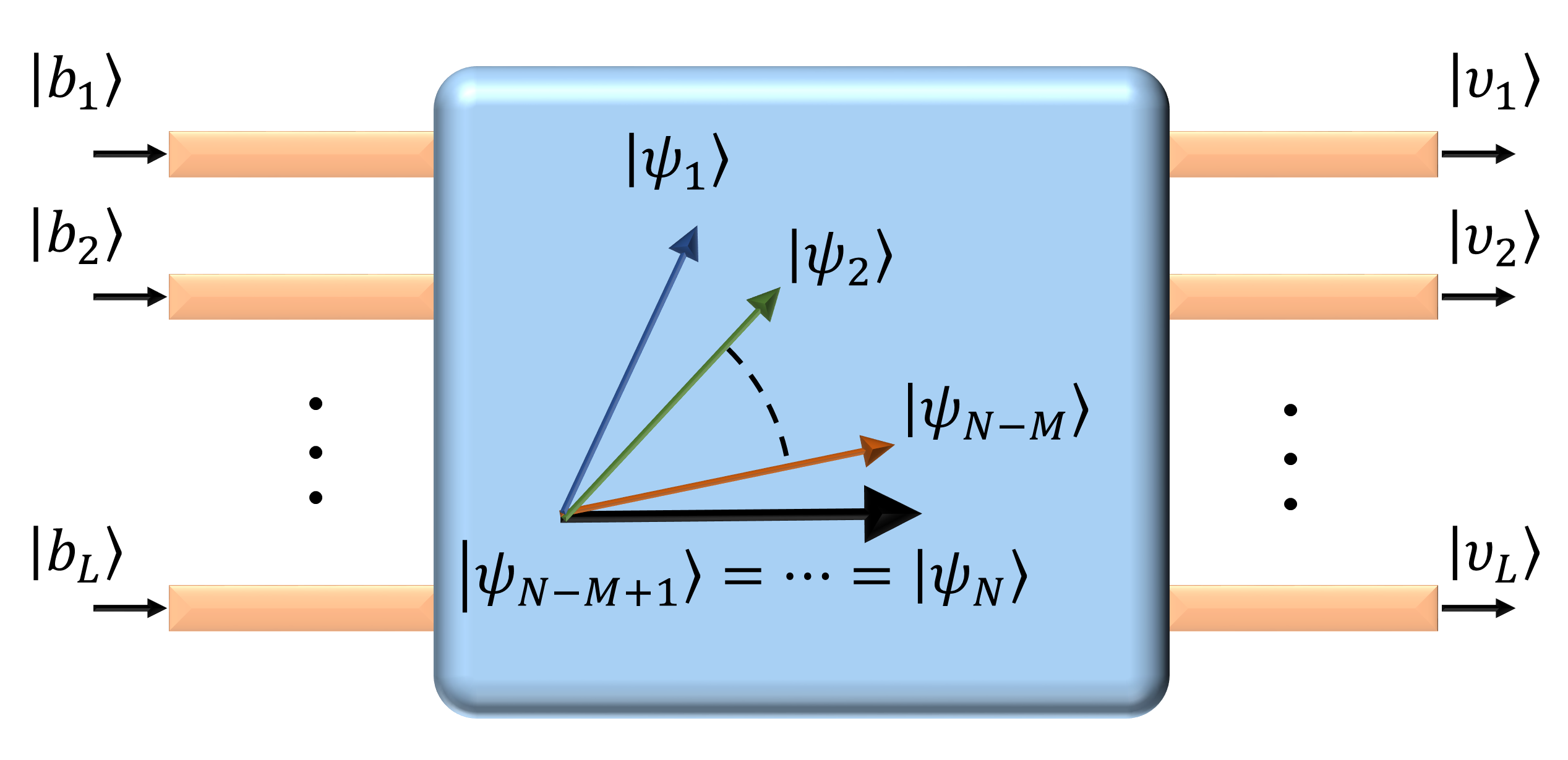}
	\caption{\textbf{A schematic of an open resonant structure with an EP and multiple input/output channels.} As indicated by the structure of the eigenvectors $\{\ket{\psi_n}\}$, the spectrum of this resonant systems is assumed to be of dimension $N$ and having an EP of order $M$ that arises due to gain/absorption or coupling to $L$ input/output ports. $\ket{b_i}$ is the input signal from $i^{th}$ input port, and $\ket{v_i}$ is the output signal from $i^{th}$ output port.}
		\label{Fig:Open_System}
\end{figure}

\textit{\textbf{Model and preparatory comments}}--- Within the context of temporal coupled mode formalism \cite{Fan2003OOP}, a complex resonant photonic structure (see Fig.\:\ref{Fig:Open_System}) under linear conditions can be modeled by the following set of equations:

\begin{equation} \label{Eq:CMT_General}
\begin{split}
i\frac{d \ket{a(t)}}{dt} &=\hat{\mathcal{H}} \ket{a(t)}+i \hat{\Gamma} \ket{b(t)} \\
\ket{v(t)} &=\hat{Y}\ket{b(t)}-\hat{\Gamma}^T\ket{a(t)},
\end{split}
\end{equation}

\noindent where the kets $\ket{a(t)}=[a_1(t),a_2(t),\ldots,a_N(t)]^T$, $\ket{b(t)}=[b_1(t),b_2(t),\ldots,b_L(t)]^T$  and $\ket{v(t)}=[v_1(t),v_2(t),\ldots,v_L(t)]^T$ represent the modal amplitudes of the resonant modes and the input and the output channels, respectively. The $N\times N$ time-independent non-Hermitian matrix Hamiltonian $\hat{\mathcal{{H}}}$ characterizes coupling between the different resonant states, whereas the $L\times L$ matrix $\hat{Y}$ quantifies the direct scattering between incoming and outgoing channels. Finally, the $N\times L$ matrix $\hat{\Gamma}$ describes the coupling between the $N$ resonant modes and the $L$ input/output channels. To simplify the notations, we also define $\ket{f(t)}\equiv i\hat{\Gamma} \ket{b(t)}$. The general solution to Eq.(\ref{Eq:CMT_General}) that takes into account the transient response can be obtained by using Laplace transform. Here, however, we are interested in the steady state response $\ket{A(\omega)}$, which can be expressed in terms of the frequency domain resolvent (sometimes also called Green's operator or function) $\hat{G}(\omega) \equiv (\omega \hat{I}-\hat{\mathcal{H}})^{-1}$ (i.e., $\hat{I}$ is the unit operator) as:

\begin{equation} \label{Eq:Fourier1}
\ket{A(\omega)}=\hat{G}(\omega) \ket{F(\omega)},
\end{equation}

\noindent where $\ket{A(\omega)} \equiv \mathcal{FT}(\ket{a(t)})$ and $\ket{F(\omega)} \equiv \mathcal{FT}(\ket{f(t)})$ with $\mathcal{FT}(\cdot)$ denoting the Fourier transform. Before we proceed, we emphasize that the existence of a finite response (i.e. non-diverging resolvent) is not always guaranteed as we will discuss in more detail later.

To keep the discussion focused, we will only consider an excitation vector $\ket{f(t)}$ with separable time dependence and spatial profile, i.e. $\ket{f(t)}=s(t) \ket{u}$ where $s(t)$ is a scalar function of time and $\ket{u}=[u_1,u_2,\ldots,u_N]^T$ is a time independent excitation profile vector. In the frequency domain, the excitation vector thus takes the form $\ket{F(\omega)}=S(\omega) \ket{u}$. The eigenvectors of the non-Hermitian $\hat{\mathcal{H}}$ defined by the set $\{\ket{\psi^r_n}: \hat{\mathcal{H}}\ket{\psi^r_n}=\Omega_n\ket{\psi^r_n}, \braket{\psi^r_n|\psi^r_n}=1\}$ have the following two important properties: (1) The eigenvalues $\Omega_n$ are in general complex; and (2) The eigenvectors $\ket{\psi^r_n}$ do not need to be orthogonal. In the absence of EPs, the eigenvectors form a complete basis, hence we can represent any input profile using the expansion  $\ket{u}=\sum_{n=1}^{N} c_n \ket{\psi_n}$, which in turn reduces Eq.(\ref{Eq:Fourier1}) to:

\begin{equation} \label{Eq:Fourier2}
\frac{\ket{A(\omega)}}{S(\omega)}=\sum_{n=1}^{N} \frac{c_n}{\omega-\Omega_n} \ket{\psi^r_n}.
\end{equation}

\noindent Note, however, that the expansion coefficients $c_n$ cannot be calculated using the usual projection $\braket{\psi_n^r|u}$. Instead, one has to employ the left eigenvectors of $\hat{\mathcal{H}}$, defined by $\{\bra{\psi^l_n}: \bra{\psi^l_n}\hat{\mathcal{H}}=\Omega_n\bra{\psi^l_n}, \braket{\psi_m^l|\psi_n^r}=\delta_{n,m}\}$, to obtain $c_n=\braket{\psi^l_n|u}$ (see supplementary note 2 for more details). This is known as bi-orthogonal projection. The alerted reader will notice that we dropped the normalization condition $\braket{\psi^l_n|\psi^l_n}=1$ from the definition of ${\bra{\psi^l_n}}$. In fact, it can be shown that, for non-normal matrices, the conditions  $\braket{\psi_m^l|\psi_n^r}=\delta_{n,m}$, $\braket{\psi_n^{r,l}|\psi_n^{r,l}}=1$ cannot be satisfied simultaneously \cite{Siegman1989PRA} (see supplementary note 3).\\
By using $c_n \ket{\psi_n^r}=\braket{\psi_n^l|u}\ket{\psi_n^r}=\ket{\psi_n^r}\braket{\psi_n^l|u}$, and by assuming that the non-orthogonal basis are complete, we can express the resolvent as \cite{Pick2017,Shomerus_2020_PRR}:

\begin{equation} \label{Eq:Fourier2}
\hat{G}(\omega)=\sum_{n=1}^{N} \frac{\ket{\psi^r_n}\bra{\psi_n^l}}{\omega-\Omega_n}.
\end{equation}

Even without considering the effect of EPs, the above results already lead to counterintuitive results. For instance, energy transfer from an excitation signal to a particular eigenmode of the system can still take place even when their Hermitian overlap vanishes. In order to illustrate this, consider an input signal $\ket{u}=\frac{\ket{\psi^r_m}}{\braket{\psi^r_n|\psi^r_m}}-\ket{\psi^r_n}$ with $\braket{\psi^r_m|\psi^r_n} \neq 0$ (due to non-Hermiticity). Clearly, $\braket{\psi^r_n|u}=0$ yet the signal $\ket{u}$ will excite the state $\ket{\psi^r_n}$. The converse is also true. An input given by $\ket{u}=\ket{\psi^r_m}$ will excite only the state $\ket{\psi^r_m}$ even though it may have a finite overlap with other states. \\

\textit{\textbf{Defective spectrum does not mean defective response}}--- We now consider the case when the spectrum of the Hamiltonian $\hat{\mathcal{H}}$ contains an EP. For generality, we assume that the EP is of order $M$, i.e. formed by the coalescence of $M$ eigenstates. Such a Hamiltonian operator is said to be \textit{defective} (i.e., the eigenstates of the Hamiltonian do not form a complete basis). In the literature of non-Hermitian optics, the response of a system described by a defective $\hat{\mathcal{H}}$ is sometimes studied using perturbative approaches \cite{Mailybaev2004} (see for instance \cite{Pick2017}). While these perturbative expansions eventually lead to correct conclusions, they complicate the analysis and give the impression that the resultant formulas are only approximations. As we describe below, this is actually not the case. On the contrary, one can use exact, non-perturbative expansions to arrive at the same final results in a straightforward fashion. In this section, we focus on the case when the system exhibits only one EP we discuss the more general case in the supplementary material.\\

To proceed, let us denote the eigenvalue and eigenvector corresponding to an EP by $\Omega_{EP}=\Omega^{R}_{EP}+i \Omega^{I}_{EP}$ and $\ket{\psi^r_{EP}}$, respectively. In order to avoid any confusion, we will also refer to the Hamiltonian and its corresponding resolvent in this case by $\hat{H}_{EP}$ and $\hat{G}_{EP}(\omega)$. Note that, since the Hamiltonian $\hat{H}_{EP}$ is defective, $\hat{G}_{EP}(\omega)$ cannot be expressed using a simple expansion similar to that of Eq.(\ref{Eq:Fourier2}). Perturbative methods such as  Newton-Puiseux series \cite{Mailybaev2004} have been used in the literature to mitigate this problem and obtain an expansion for $\hat{G}_{EP}(\omega)$. What is often overlooked, however is that even for a defective $\hat{H}_{EP}$, the operator $(\omega \hat{I}-\hat{H}_{EP})$ can still be invertible. This key observation can tremendously simplify the analysis in many situations. Particularly, when an optical resonator system operates in the passive mode (i.e., no optical gain) or even in active mode below the lasing threshold (i.e., as in amplifiers), the complex eigenvalue lies in the lower half of the complex plane, i.e. $\Omega^I_{EP} < 0$. On the other hand, probing the response of such a system is done by using external excitation that has a real frequency, i.e. along the real axes. In this case, the corresponding eigenvalue of $\hat{G}_{EP}(\omega)$, which is given by $\mu_{EP}=\left[\omega-(\Omega^R_{EP}+i\Omega^I_{EP})\right]^{-1}$, does not diverge even at resonance when $\omega=\Omega^R_{EP}$. As a result, one can evaluate $\hat{G}_{EP}(\omega)$ by direct matrix inversion for any value of the real frequency $\omega$. In principle, this information is sufficient for characterizing the linear response of the resonator, but it does not provide much insight into the interplay between the excitation profile and the response. In addition, one may wonder about the fate of the expansion in (\ref{Eq:Fourier2}). As we mentioned earlier, Newton-Puiseux series is often used to generalize this expression. We now show that this generalization is in fact exact and does not employ any perturbation analysis. To do so, we first note that the nondegenerate eigenvectors of an $N \times N$ matrix $\hat{G}_{EP}$ that has an EP of order $M$ span only a reduced $N-M$ dimensional space, call it $\mathcal{D}_{\phi}$. We will denote the missing domain by $\mathcal{D}_{EP}$. In order to form a complete basis, we follow the standard  Jordan chain procedure \cite{Igor2013book}, defined by the set of vectors that satisfy the following recursive equations:

\begin{equation} \label{Eq:JordanChain}
	\begin{aligned}
		(\hat{H}_{EP}-\Omega_{EP}\hat{I})\ket{J_1^r} &=0  \\
		(\hat{H}_{EP}-\Omega_{EP}\hat{I})\ket{J_2^r} &=\chi_2 \ket{J_1^r} \\
		 &\vdots \\
		(\hat{H}_{EP}-\Omega_{EP}\hat{I})\ket{J_M^r} &=\chi_{M} \ket{J_{M-1}^r}.
	\end{aligned}
\end{equation}

In the above, we used the notation $\ket{J_1^r}\equiv\ket{\psi_{EP}^r}$ for clarity. The constants $\chi's$ are introduced to ensure the consistency of the physical dimensions, and their values are chosen to achieve normalization, i.e. $\braket{J_n^r|J_n^r}=1$ for any integer $n=1,2,\ldots, M$. The vectors $\ket{J_n^r}$ are generalized right eigenvectors of the operator $\hat{H}_{EP}$, i.e. they satisfy the eigenvalue problem $(\hat{H}_{EP}-\Omega_{EP}\hat{I})^n\ket{J_n^r}=0$ which implies that $\braket{\psi_m^l|J_n^r}=0$ for any $n$ and $\bra{\psi_m^l}\neq\bra{\psi_{EP}^l}$ (see supplementary note 2). Since the vectors $\ket{J_n^r}$ are linearly independent by construction (see supplementary note 2 for a brief proof), it follows that they span the domain $\mathcal{D}_{EP}$, and thus complete the basis. Note however that while $\ket{J_1^r}$ is unique (up to a constant), there is a freedom in choosing the set of other vectors $\ket{J_n^r}$ for $n>1$. Intuitively, this situation is similar to fixing the $z$ axis in three dimensions and rotating the $x$ and $y$ axes in the $x-y$ plane around the origin. Thus, in general extra normalization conditions are required in order to fix the choice of the vectors $\ket{J_n^r}$. Here, we will not be concerned with the exact orientation of $\ket{J_n^r}$. Another important observation is that some of the $\ket{J_n^r}$ vectors are self-orthogonal (see supplementary note 2). Thus, while any arbitrary input can be decomposed according to $\ket{u}=\sum_{n=1}^{N-M} c_n \ket{\psi_n^r}+\sum_{m=1}^{M} d_m \ket{J_m^r}$ with the constants $c_n=\braket{\psi_n^l|u}$ evaluated using biorthogonality as before, the coefficients $d_m$ cannot be obtained directly using the same strategy. However, as we show in supplementary note 2, one can use the vectors of ($\bra{J_m^l}$) to define another set of vectors ($\bra{\tilde{J}_m^l}$) that satisfy the relation $\braket{\tilde{J}_m^l|J_n^r}=\delta_{m,n}$ and hence $d_m=\braket{\tilde{J}_m^l|u}$.\\

\begin{figure}
	\includegraphics[width=5in]{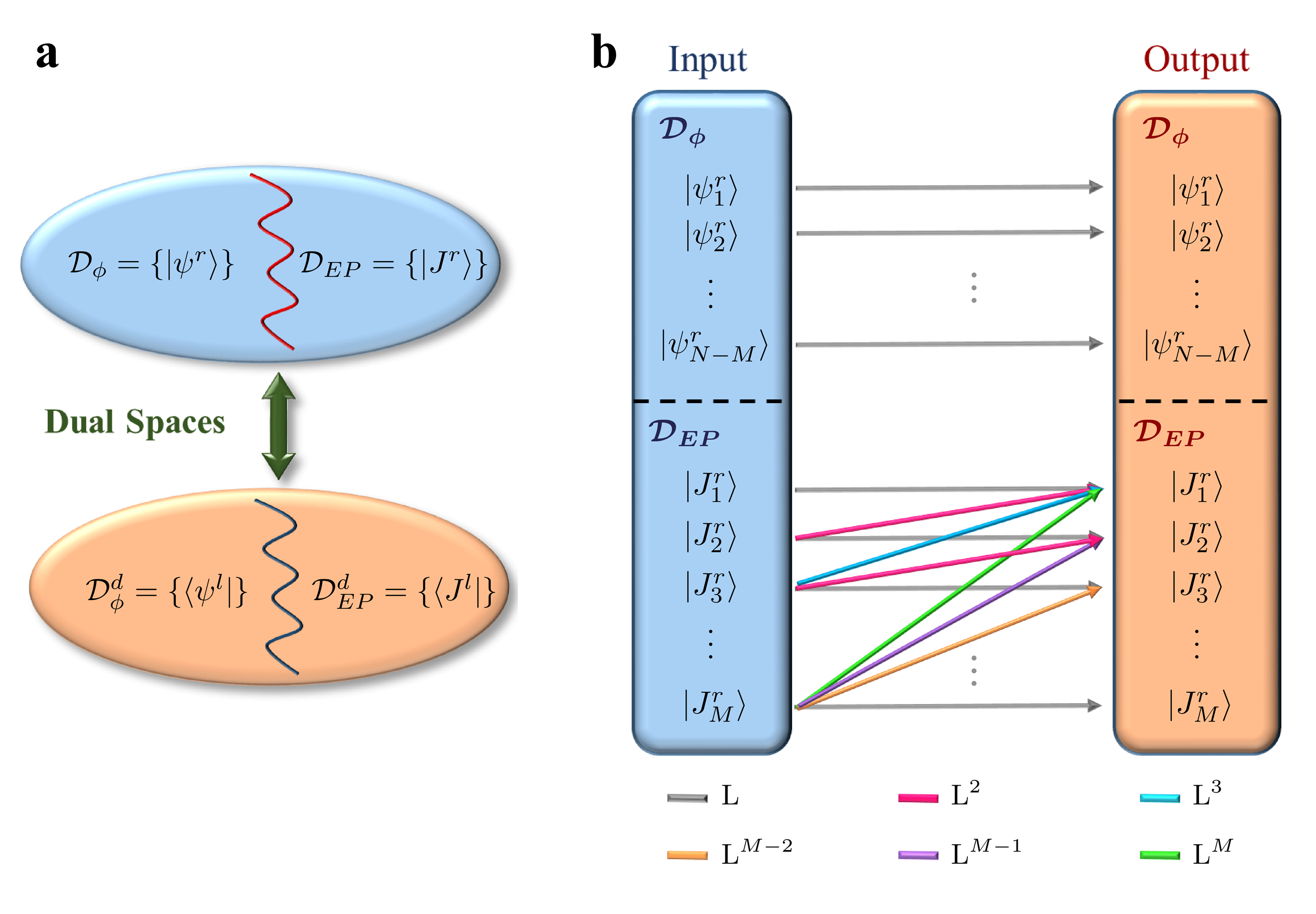}
	\caption{\textbf{Structure of the eigenspace and its fingerprint on the linear response.} \textbf{a} The underlying vector space associated with the non-Hermitian Hamiltonian $\hat{\mathcal{H}}$ can be divided into two subspaces: $\mathcal{D}_{\phi}=\{\ket{\psi_n^r}:n=1,2,\ldots,N-M\}$ and $\mathcal{D}_{EP}=\{\ket{J_m^r}: m=1,2,\ldots,M\}$. The former is spanned by the non-degenerate right eigenvectors of $\hat{\mathcal{H}}$, while the latter is spanned by the right generalized eigenvectors. Note that in this classification, the degenerate (or exceptional) vector belongs to $\mathcal{D}_{EP}$. The dual spaces $\mathcal{D}_{\phi}^d$ and $\mathcal{D}_{EP}^d$ are defined in a similar fashion for the left ordinary and generalized eigenvectors. \textbf{b} Pictorial representation of the linear response associated with a non-Hermitian system having an EP. Typical Lorentzian response arises due to coupling between an input and an output channel that belong to the same modal class. On the other hand, super-Lorentzian responses emerge when the input signal matches a particular generalized eigenmodes of certain order while the output signal matches a lower order generalized eigenvector (including the exceptional vector) according to Eq.\:\ref{Eq:Greens1}. The arrows, together with the legends, illustrate a few possible responses explicitly. The symbol $\text{L}^m$ indicates a super-Lorentzian response of order $m$, i.e. a Lorentzian raised to the power $m$.}
	\label{Fig:Dual_Space}
\end{figure}

Next, given the above input signal $\ket{u}$, we seek a similar expansion of the output, i.e. in the form: $\hat{G}_{EP} \ket{u}=\sum_{n=1}^{N-M} \tilde{c}_n\ket{\psi_n^r}+\sum_{m=1}^{M} \tilde{d}_m\ket{J_m^r}$, or equivalently \\
 $(\omega \hat{I}-\hat{H}_{EP})\left\{\sum_{n=1}^{N-M} \tilde{c}_n\ket{\psi_n^r}+\sum_{m=1}^{M} \tilde{d}_m\ket{J_m^r}\right\}= \sum_{n=1}^{N-M} c_n \ket{\psi_n^r}+\sum_{m=1}^{M} d_m \ket{J_m^r}$. By applying the operator $\omega \hat{I}-\hat{H}_{EP}$ to each term inside the summation and rearranging, we obtain a sum of the form $\sum_{n=1}^{N-M} C_n\ket{\psi_n^r}+\sum_{m=1}^{M} D_m\ket{J_m^r}=0$. By noting that all the vectors $\ket{\psi_n^r}$ and $\ket{J_m^r}$ are linearly independent, we find that the above relation can be satisfied if and only if all the coefficients $C_n=D_m=0$ for every $n$ and $m$. This finally leads to the expansion (see supplementary note 4 for details):

\begin{align}\label{Eq:Greens1}
\hat{G}_{EP}=\sum_{n=1}^{N-M}\frac{\ket{\psi^r_n}\bra{\psi_n^l}}{\omega-\Omega_n}+
\sum_{m=1}^{M}\sum_{k=m}^{M}\alpha_k^{(m)}\frac{\ket{J^r_m}\bra{\tilde{J}^l_k}}{(\omega-\Omega_{EP})^{k-m+1}},
\end{align}

\noindent where the coefficients $\alpha$'s are defined by the recursive relations $\alpha_m^{(m)}=1,\ \alpha_{k}^{(m)}=\alpha_{k-1}^{(m)}\chi_{k}$ for $m<k\leq M$ (see supplementary note 4).

 We now pause to make several comments on the above expression. Firstly, the expansion series is finite and hence no convergence analysis is needed. Secondly, due to the same reason (finite terms in the series), its completeness is guaranteed. Importantly, the above expression is exact and non-perturbative despite the fact that the spectrum of $\mathcal{\hat{H}}$ contains an EP. Another important observation is that the expansion in Eq.\:(\ref{Eq:Greens1}) is not unique: a different choice of the vectors $\ket{J_m^r}$ with $m>1$ will lead to a different series (though more complicated one). Thirdly, unlike previous works that considered small systems \cite{SunadaPRA2018}, expressions for the resolvent that applies only in the vicinity of EPs \cite{Heiss2015IJTP} or expansions of the resolvent as a power series of the Hamiltonian itself \cite{Kato1995Book}, the above expansion is valid everywhere in the frequency domain for any system with arbitrary size and is expressed in terms of the eigenvector and canonical vectors. Finally, we emphasize a crucial point: expression (\ref{Eq:Greens1}) for the resolvent is evaluated for a system ``with" an EP not ``at" an EP. This semantic difference has caused confusion in the literature, mainly conveying an impression that these systems can be studied only within the context of perturbation analysis. An EP is a characteristic of the system itself, not the probe. The former can contain an EP in its spectrum that lies in the complex plane away from the real axes and still be probed with a signal that has a real frequency to obtain finite response without any singularities or divergences. On the other hand, probing the system at an EP entails either using a probe with complex frequency \cite{Baranov2017PMK,Trainiti2019OOP,Qi2020PRR} or supplying enough gain to bring the complex exceptional eigenvalue to the real axis. In both cases, the resolvent will diverge, which corresponds to the fact that the amplitude of the oscillations will grow indefinitely. In reality however, this does not happen because nonlinear effects regulate the dynamics (think of gain saturation in laser systems for instance). \\

\textit{\textbf{Super-Lorentzian frequency response}}---
In realistic configurations, a resonant structure interacts with its environment via certain scattering or coupling channels defined by the geometry. One can either probe the system via individual channels or by excitation profiles that are superpositions of several channels. At the abstract level, the concepts of channel and excitation profile are the same since they are just related by unitary transformations. In this section, we will focus on channels that directly excite the vectors $\ket{\psi_n^r}$ and $\ket{J_m^r}$ and analyze their scattering/coupling characteristics. We begin by rewriting the expression in Eq.(\ref{Eq:Greens1}) after unfolding the double summation:

\begin{equation} \label{Eq:Greens2}
\hat{G}_{EP}(\omega) = \sum_{n=1}^{N-M} \frac{\ket{\psi^r_n}\bra{\psi_n^l}}{\omega-\Omega_n} 
+\sum_{k=1}^{M} \frac{\alpha_k^{(1)}\ket{J_1^r}\bra{\tilde{J}_k^l}}{(\omega-\Omega_{EP})^k} +
\sum_{k=2}^{M} \frac{\alpha_k^{(2)}\ket{J_2^r}\bra{\tilde{J}_k^l}}{(\omega-\Omega_{EP})^{k-1}} + 
\cdots
+\frac{\ket{J_M^r}\bra{\tilde{J}_M^l}}{\omega-\Omega_{EP}}.
\end{equation}

Clearly, an input signal with a profile $\ket{u}=\ket{\psi_n^r}$ will only excite the mode $\ket{\psi_n^r}$ with a Lorentzian response centered at $\Omega_n$ as expected. Similarly, an input that matches the exceptional vector $\ket{u}=\ket{J_1^r}$ will excite only the mode $\ket{J_1^r}$, also with a Lorentzian response. A more interesting situation arises when the excitation profile coincides with a higher order Jordan vector, say $\ket{u}=\ket{J_k^r}$. In that case, according to Eq.\:(\ref{Eq:Greens2}), the exceptional eigenvector will be excited with a frequency response that features an $k^{th}$ order super-Lorentzian lineshape (i.e. a Lorentzian function raised to the power $k$). Additionally, each of the states $\ket{J_m^r}$ with $1<m \leq k$ (which are not eigenstates of $\hat{H}_{EP}$) will be also excited with a frequency response that corresponds to a super-Lorentzian of order $k-m+1$. At this point, it is important to reiterate our previous comment on the freedom of choosing the set $\ket{J_m^r}$. While $\ket{J_1^r}$ is unique, the vectors $\ket{J_m^r}$, $1<m\leq M$ are not. The consequences of this observation are not trivial. For instance, in order to excite the exceptional vector with a certain super-Lorentzian response, one can choose from a continuous manifold of excitation profiles.  To illustrate this, we consider the excitation of $\ket{J_1^r}$ with a second order super-Lorentzian response. This can be done by using the input $\ket{u}=\ket{J_2^r}+x\ket{J_1^r}$, where $x$ is a free parameter. We anticipate that these general results, which are illustrated schematically in Fig. \ref{Fig:Dual_Space}\textbf{b}, will be instrumental in developing a more comprehensive understanding of the quantum limits of several non-Hermitian optical devices (such as lasers, amplifiers, and sensors) operating at EPs. Equally important, the above analysis provides a complete picture of how the frequency response of non-Hermitian systems with EPs scales as a function of the input profile, which can be of great utility in engineering devices that rely on this feature, such as optical amplifiers with relaxed gain-bandwidth restrictions \cite{Metelmann2014PRL,Qi2020PRApplied}. In addition, our analysis also can be used to classify the input channel according to their interference effects inside the resonant system as we explain in details in supplementary note 5 and 6. Finally, we note that, so far we have focused on situations where the system has only one EP. Supplementary note 7 discusses the case when the spectrum of the relevant Hamiltonian has multiple EPs.\\

\textit{\textbf{Connection with experiments}}---Here we present illustrative realistic examples that demonstrate some of the results discussed in this work as well as the power and insight provided by our formalism which expresses the resolvent operator as an exact expansion series of the right/left canonical vectors.\\

\begin{figure}
	\includegraphics[width=5.3in]{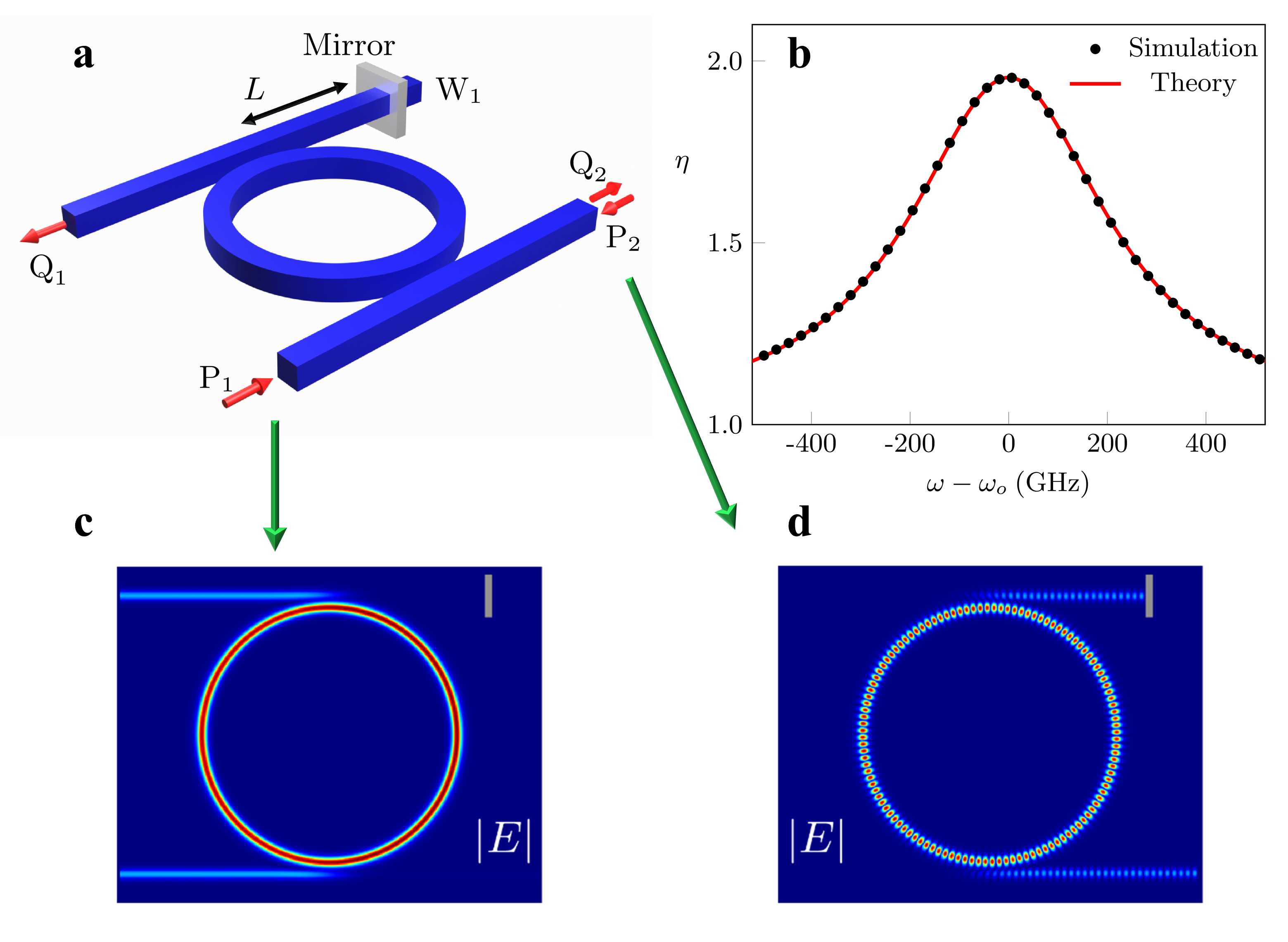}
	\caption{\textbf{Optical energy inside a photonic resonator with a second order EP}. \textbf{a} A schematic of a photonic structure that exhibits a second order EP: it consists of a microring resonator evanescently coupled to two identical waveguides, one of which is terminated with a mirror at port $\text{W}_1$. The exceptional eigenstate $\ket{J_1^r}=\ket{\psi_{EP}^r}$ can be excited through port $\text{P}_1$, while the generalized eigenstate $\ket{J_2^r}$ can be excited via port $\text{P}_2$. \textbf{b} Plot of $\eta$ (enhancement factor of stored energy in the microring) when the system is excited by an input that matches $\ket{J_2^r}$ compared to an excitation matching $\ket{\psi_{EP}}$ as a function of frequency detuning near resonance as obtained by full-wave simulations (black dots) and the closed form expression (red line). \textbf{c} and \textbf{d} Distributions of the electric field amplitudes under excitation either from ports $\text{P}_1$ or $\text{P}_2$ are plotted. As expected from the analysis, the case when $\ket{u_1}=\ket{\psi_{EP}^r}$ leads to energy storage in both the CW and CCW modes as evidenced by the standing wave pattern in \textbf{d}. All simulations were performed by using the finite element method available from the COMSOL software package. The optical parameters of the structure and the simulations details are discussed in supplementary note 8.}
	\label{Fig_setup}
\end{figure}

Our first example is depicted in Fig.\:\ref{Fig_setup}\textbf{a}. It consists of a microring resonator evanescently coupled to two identical waveguides, with one port of, say, $\text{W}_1$ terminated by a mirror. This geometry was introduced in \cite{Zhong2019SES} and shown to exhibit an exceptional surface with potential applications for sensing and controlling spontaneous emission \cite{Amin2020PRR}. In addition, it has been recently implemented in microtoroid resonators with the feedback realized using a fiber loop mirror \cite{soleymani2021chiral}. It is described by the following set of equations:

\begin{align} \label{Eq:CMT_chiral}
i \frac{d}{dt}
\left[\begin{matrix}
a_1(t) \\
a_2(t)\\
\end{matrix}\right]
&=\left[
\begin{matrix}
\omega_o-2i\gamma & 0 \\
\kappa & \omega_o-2i\gamma\\
\end{matrix}\right]
\left[\begin{matrix}
a_1(t)\\
a_2(t)
\end{matrix}\right]
+ i \sqrt{2\gamma}
\left[
\begin{matrix}
b_1(t) \\
b_2(t)\\
\end{matrix}\right]\nonumber\\
Q_1&=-\sqrt{2\gamma}\ a_2\nonumber\\
Q_2&=P_1-\sqrt{2\gamma}\ a_2,
\end{align}

\noindent where, $\omega_o$ is the resonant frequency, $\gamma$ is the decay rate of the resonant mode into each waveguide. In addition, $\kappa=-2i \gamma |r| e^{i\phi}$, where $|r|$ is the absolute value of the mirror amplitude reflection coefficient, and $\phi\equiv 2\beta L + \phi_r$ is a phase factor that quantifies the phase of the mirror reflectivity, $\phi_r$, and its distance from the resonator, $L$, where $\beta$ is the propagation constant in the waveguide.\\

The resolvent in this case, which we will denote by $G_2$ can be evaluated in closed form and is given by:

\begin{equation} \label{Eq:Chiral_Green_operator}
G_2(\omega)= \left[
\begin{matrix}
\frac{1}{\Delta \omega+2i\gamma} & 0 \\
 & \\
\frac{\kappa}{\left(\Delta \omega+2i\gamma\right)^2} & \frac{1}{\Delta \omega+2i\gamma}
\end{matrix}\right],
\end{equation}

\noindent where $\Delta \omega=\omega-\omega_o$. It is straightforward to show that $\ket{\psi_{EP}^r}\equiv \ket{J_1^r}=[0 , 1]^T$ and $\ket{J_2^r}=\frac{\chi}{\kappa} [1,0]^T$. Note that $\braket{J_1^r|J_1^r}= 1$, while $\braket{J_2^r|J_2^r}\neq 1$, i.e. the latter is not normalized. If we now consider the two normalized inputs $\ket{b_{1,2}}=\alpha_{1,2} \ket{J_{1,2}^r} e^{-i\omega t}$ with $\alpha_1=1$ and $\alpha_2=\frac{\kappa}{\chi}$ we find that the normalized stored energies are given by $\braket{E_1}=\frac{2\gamma}{\Delta \omega^2 +4\gamma^2}$ and  $\braket{E_2}=\frac{2\gamma}{\Delta \omega^2 +4\gamma^2} \left[1+\frac{4\gamma^2|r|^2}{\Delta \omega^2 +4\gamma^2}\right]$. Thus, $\eta\equiv \frac{\braket{E_2}}{\braket{E_1}}=1+\frac{4\gamma^2|r|^2}{\Delta \omega^2 +4\gamma^2}>1$. At resonance when $\Delta \omega=0$, we find that $\eta=2$ when a completely reflecting mirror, $r=1$, is used.

\begin{figure}
\includegraphics[width=5.5in]{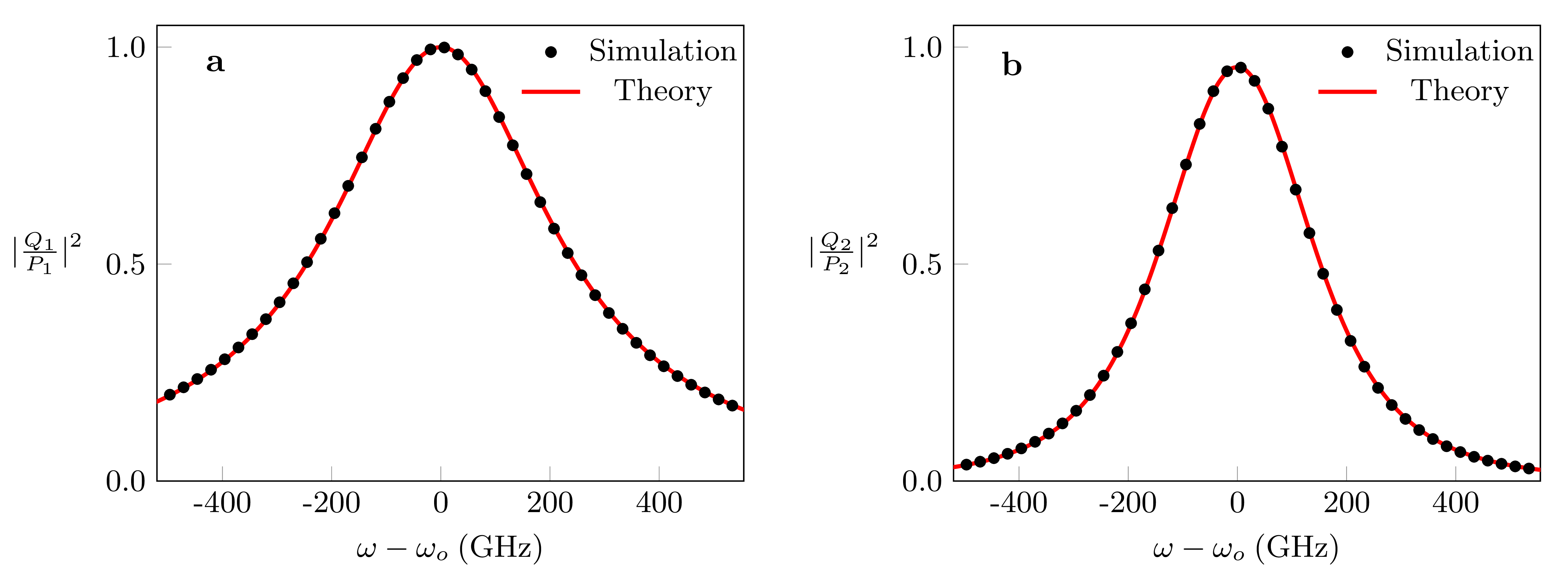}
\caption{\textbf{Illustration of the Lorentzian and square-Lorentzian responses of the system. a} An input signal is launched form port $\text{P}_1$ (see Fig. \ref{Fig_setup}\textbf{a}) and the normalized output power $|\frac{Q_1}{P_1}|^2$ with Lorentzian distribution is observed at port $\text{Q}_1$. \textbf{b} For an input signal from port $\text{P}_2$ a square-Lorentzian response for the normalized output power $|\frac{Q_2}{P_2}|^2$ is collected at port $\text{Q}_2$. Red solid line shows the theoretical results while black dots show results obtained through simulations.}
\label{Fig:square_lorentzian}
\end{figure}

To confirm these predictions, we consider a realistic photonic implementation of the structure of Fig.\:\ref{Fig_setup}\textbf{a} as explained in detail in supplementary note 8. It is straightforward to show that excitations from ports $\text{P}_{1,2}$ are mode-matched with $\ket{J_{1,2}^r}$, respectively. Figure \ref{Fig_setup}\textbf{b} plots the value of $\eta$ as a function of the frequency detuning as obtained from the full-wave simulations (black dots) as well as by direct substitution in the closed form expression for $\eta$ described above (red line). On the other hand, Figs. \ref{Fig_setup}\textbf{c} and \textbf{d} depict the steady state field distribution as obtained by full-wave analysis (see supplementary note 8 for the details of the optical parameters of the structure used in the simulations) inside the system under the two different excitations at resonance. In order to study the spectral response of this configuration, we plot the scattering coefficients $|\frac{Q_1}{P_1}|^2$ and $|\frac{Q_2}{P_2}|^2$ in Figs. \ref{Fig:square_lorentzian}\textbf{a} and \textbf{b}, where the black dots represent data obtained from full wave simulations and red line indicate theoretical results. As predicted by our analysis [see Eq. (\ref{Eq:CMT_chiral})], $|\frac{Q_1}{P_1}|^2$ which corresponds to an excitation and collection from the exceptional vector features a Lorentzian response, while $|\frac{Q_2}{P_2}|^2$ which corresponds to an excitation matching $\ket{J_2^r}$ and collection matching $\ket{\psi_{EP}^r}$ follows a super-Lorentzian of order two. This simple but intuitive example demonstrate a subtle  property of open systems, namely that their response lineshape is not unique but rather depends on the input/output channel configuration. Not only this feature arises naturally from our analysis, but the exact expansion series for resolvent operator [see Eq. (\ref{Eq:Greens1})] allows us to tailor this response at will by selecting the appropriate coupling channels. Importantly, we stress that this scenario is different from the previous works in which a control parameter, such as coupling strength, frequency detuning, or loss imbalance between resonance modes, is tuned to feature a Fano, EIT, or an ATS lineshape \cite{LiAppPhysLett2011, PengNatComm2014, WangIEEEAccess2020, MaJLightwTechnol2021}. In contrast, for the system studied here, there is no such control parameter: one can obtain a Lorentzian or a squared-Lorentzian lineshape by choosing the appropriate input and output channel pairs.

\begin{figure}
\includegraphics[width=5.3in]{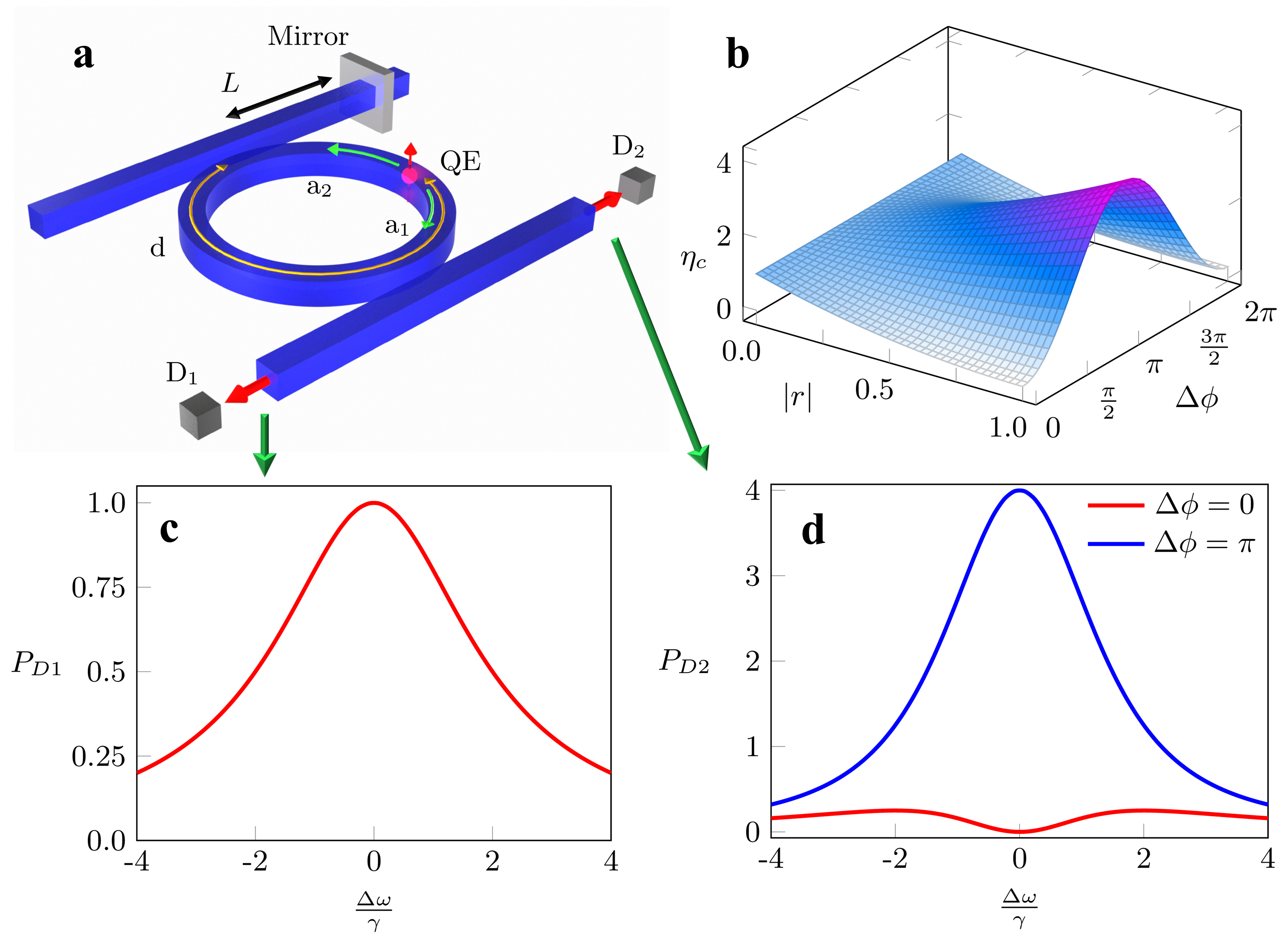}
\caption{\textbf{Spontaneous emission at exceptional surfaces.} \textbf{a} A photonic system consisting of a quantum dot embedded in a microring resonator evanescently coupled to a waveguide with an end mirror that induces a unidirectional coupling between these modes. The emission from the quantum dot couples to the CW and CCW modes ($a_{1,2}$) of the ring and eventually detected by the detectors $D_{1,2}$. \textbf{b} The ratio $\eta_c$ of the stored energy in the CCW mode to the stored energy in the CW wave is plotted for the resonant frequency $\omega=\omega_o$. \textbf{c} The normalized power detected by $D_1$ has a Lorentzian distribution. \textbf{d} The normalized power detected by $D_2$ for $\Delta\phi=0$ (red curve) and $\Delta\phi=\pi$ (blue curve) are shown for $|r|=1$. As explained in the text, the detected power spectrum at $D_2$ features an interference between a Lorentzian and a square-Lorentzian terms, with the final outcome strongly depending on the relative position between the mirror and the quantum dot as quantified by the parameter $\Delta \phi$.}
\label{Fig:Emitter_Resonator}
\end{figure}

Our analysis extends beyond the simple input/output evanescently coupled channels and can be useful also in understanding light-matter interaction at EPs. As an example, consider the geometry depicted in Fig. \ref{Fig:Emitter_Resonator}\textbf{a} which consists of a quantum dot embedded inside a microring-waveguide-mirror system implementing an exceptional surface. This system was investigated recently and it was shown that it can be used to control the rate of spontaneous emission and Purcell factor \cite{Amin2020PRR, Khanbekyan2020PhysRevResearchDSS, ren2021quasinormal}. Let us now assume that two single-photon detectors, $D_{1,2}$ are located as shown in the figure. What would be the spectrum of the detected photon as measured by each detector after repeating the experiment a large number of times? This question can be answered only by calculating the Green's operator matrix elements. While this task can be done easily by a matrix inversion in this simple case (see \cite{Amin2020PRR} for more details), this approach does not give any insight into the physics of the system. On the other hand, by applying our formalism, It is straightforward to see, based on the photon trajectories that detector $D_1$ will report a Lorentzian response while $D_2$ will report a superposition of Lorentzian and super-Lorentzian of order two. More rigorously, the system here is described by equations similar to that presented in Eq. (\ref{Eq:CMT_chiral}) (see \cite{Amin2020PRR} for details), where here $\ket{b_{1,2}}$ represent the emission from the quantum dot and are given by:
$\left[b_1(t), b_2(t)\right]^T=\frac{J_e e^{-i\omega_et}}{i\sqrt{2\gamma}}\left[e^{-i\phi_E}, e^{i\phi_E}\right]^T$, where $\omega_e$ is the transition frequency of the QE, the phase $\phi_E=\beta d$, and $\beta$ is the propagation constant of the modes (see \cite{Amin2020PRR} for a detailed derivation). By using the resolvent in Eq. (\ref{Eq:Chiral_Green_operator}), the response of the system is calculated as: $G_2(\omega_e)[J_e e^{-i\phi_E},J_e e^{i\phi_E}]^T=\frac{J_e}{\Delta\omega+2i\gamma}[e^{-i\phi_E},e^{i\phi_E}+\frac{\kappa}{\Delta\omega+2i\gamma}e^{-i\phi_E}]^T$. The above expression indicates that the power spectrum of CW mode associated with the microring structure in Fig. \ref{Fig:Emitter_Resonator}\textbf{a} is Lorentzian while that of the CCW mode features a superposition of Lorentzian and square-Lorentzian terms. In addition, by recalling that $\kappa$ is a function of the mirror reflectivity and location, it is straightforward to see that the interference effects associated with the mixed term will vary as a function of the optical path between the mirror and the emitter $\Delta\phi\equiv 2\phi_E-\phi=2\beta(d-L)+\phi_r$. Figure \ref{Fig:Emitter_Resonator}\textbf{b} depicts the ratio of the stored energy in the CCW mode to that in the CW mode  $\eta_{c}\equiv \frac{\braket{E_2}}{\braket{E_1}}=|1-\frac{2i\gamma|r|e^{-i\Delta\phi}}{\Delta\omega+2i\gamma}|^2$ as a function of the mirror reflectivity $|r|$ and $\Delta \phi$, which clearly highlights how the response of the intra-cavity modes depends on the relative position of the mirror with respect to the emitter. For completeness, we also plot the normalized output power spectra $P_{D1,D2}=|\frac{\sqrt{2\gamma}\:Q_{1,2}}{J_e}|^2$ detected by $D_{1,2}$ as a function of the normalized frequency detuning $\frac{\Delta\omega}{\gamma}$ as shown in Figs \ref{Fig:Emitter_Resonator}\textbf{c} and \textbf{d}, respectively. As expected, $P_{D1}=\frac{4\gamma^2}{\Delta\omega^2+4\gamma^2}$ is Lorentzian, while $P_{D2}=\frac{4\gamma^2}{\Delta\omega^2+4\gamma^2}|1-\frac{2i\gamma|r|e^{-i\Delta\phi}}{\Delta\omega+2i\gamma}|^2$ features an interference between a Lorentzian and square-Lorentzian terms. While the above example considers realistic scenarios of a quantum dot having a small size compared to the wavelength of light, the notion of controlling spontaneous emission using exceptional points could be also relevant to other setups involving artificial atoms with larger size where the dipole approximation fails \cite{KannanNat2020}, as well as systems consisting of multiple quantum emitters \cite{Ganainy_NJP2013}.\\

\begin{figure}
\includegraphics[width=6.8in]{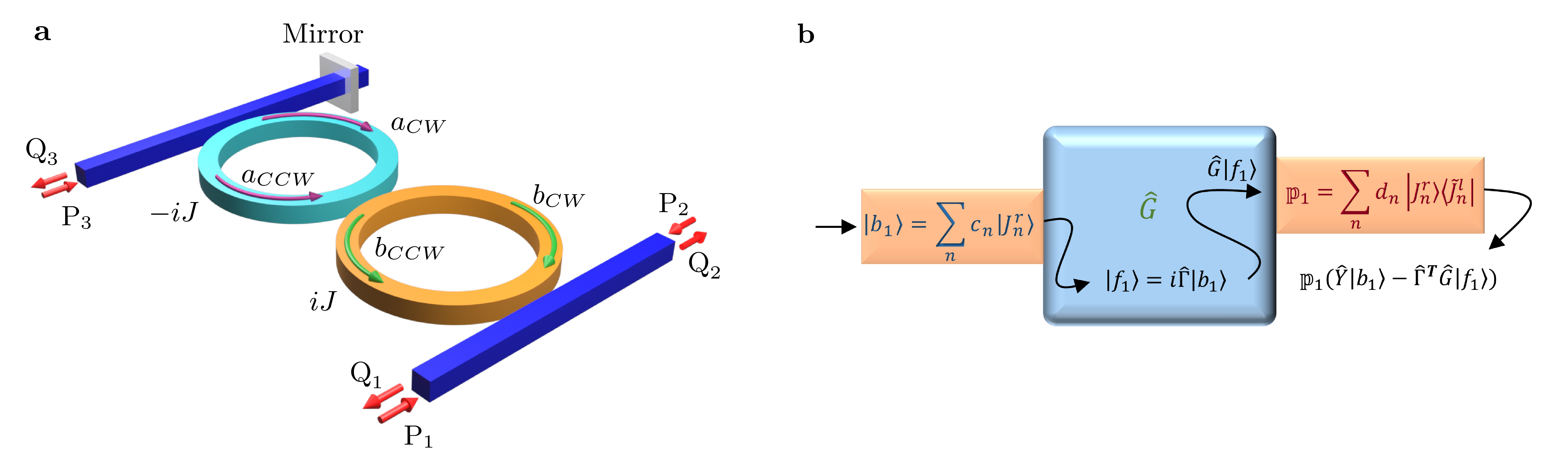}
\caption{\textbf{A PT symmetric system with unidirectional coupling between the CW/CCW modes.} \textbf{a} A schematic of the studied system, which was introduced in \cite{Zhong2020PhysRevLett.125.203602} and shown to exhibit an EP of order 4. $J$ is the gain/loss factor. $P_{1,2,3}$ are input signals and $Q_{1,2,3}$ are output signals. $a_{CW,CCW}$ and $b_{CW,CCW}$ are CW and CCW modes of the cavities. \textbf{b} Pictorial depiction of the systematic approach to obtain the system's linear response using the formalism presented in this work. $\ket{b_1}$ is the excitation profile expanded with respect to the right eigenvectors $\{\ket{J_n^r}\}$. $\{\bra{\tilde{J}_n^l}\}$ are the corresponding left bi-orthogonal vectors. $\hat{G}$ is the resolvent and $\mathbb{P}_1$ is the projection operator assigned to the output channel. $\hat{Y}$ quantifies the direct scattering between incoming and outgoing channels and $\hat{\Gamma}$ describes the coupling between the resonant modes and the input/output channels.}
\label{Fig:EP4_setup}
\end{figure}

Having demonstrated the utility of our approach and the insight it provides using a simple system with one resonant element, we now consider a more complex system that consists of a PT symmetric arrangement and an asymmetric delayed feedback as shown in Fig. \ref{Fig:EP4_setup}\textbf{a}. In the absence of the mirror, this structure exhibits two independent EPs, each of which of order two. The introduction of the mirror causes these two EPs to coalesce, forming an EP of order four \cite{Zhong2020PhysRevLett.125.203602}. Despite the fact that this system consists of only two ring resonators, the presence of inhomogeneous gain and loss distribution makes it difficult to trust intuition in this case. On the other hand, brute-force analysis by evaluating the scattering coefficients one at a time does not provide much insight into the scattering profile. To analyze this system using our formalism, we first consider the Hamiltonian matrix describing this system as written in the natural basis of the individual, isolated ring resonators $[a_{CW},b_{CCW},a_{CCW},b_{CW}]^T$:
\begin{equation}
\hat{H}_4=\left[
\begin{smallmatrix}
 \omega _o-i \gamma -i J & J & 0 & 0 \\
 J & \omega _o-i \gamma +i J & 0 & 0 \\
 \kappa & 0 & \omega _o-i \gamma -i J & J \\
 0 & 0 & J & \omega _o-i \gamma +i J \\
\end{smallmatrix}
\right],
\end{equation}
 Here, $\omega_o$ is the resonant frequency, $\gamma$ is the decay rate of the resonant mode into each waveguide, $J$ is the coupling rate between the two rings and also the gain/loss factors in the yellow/green rings respectively, i.e. the system respects PT symmetry. In addition, $\kappa=-2i\gamma|r|e^{i\phi}$, where $|r|$ is the absolute value of the mirror reflection coefficient, and the phase factor $\phi$ quantified the phase of the mirror reflection coefficient and its distance from the adjacent resonator. The right (canonical) eigenvectors associated with this Hamiltonian are given  $\ket{J_1^r}=[0,0,-i,1]^T$, $\ket{J_2^r}=[0,0,1,0]^T$, $\ket{J_3^r}=[1,i,0,0]^T$, and $\ket{J_4^r}=[0,1,0,0]^T$. These vectors are calculated based on the following choice for constant coefficients $\chi's$ [see Eq. (\ref{Eq:JordanChain})]: $\chi_2=\chi_4=J,\ \chi_3=\kappa$. The corresponding  left bi-orthogonal vectors, which satisfy the relation $\braket{\tilde{J}^l_n|J^r_m}=\delta_{nm}$, are then given by: $\bra{\tilde{J}_1^l}=[-i,1,0,0]$, $\bra{\tilde{J}_2^l}=[1,0,0,0]$, $\bra{\tilde{J}_3^l}=[1,i,0,0]$, and $\bra{\tilde{J}_4^l}=[0,0,0,1]$. The scattering profile and spectral response of the system can be then evaluated by using  Eq. (\ref{Eq:Greens1}) and using the following systematic steps: (1) Write down the excitation profile in the basis of the bare eigenmodes $a_{CW}$, $a_{CCW}$, $b_{CW}$, and $b_{CCW}$;  (2) Express this vector in the basis of the right (canonical) eigenvectors;  (3) Obtain the response using Eq. (\ref{Eq:Greens1}); (4) Assign a projection operator to each output channel. For instance, a unit input signal from port $P_1$ will directly couple only to the field amplitude $b_{CCW}$(see Fig. \ref{Fig:EP4_setup}\textbf{a}), and hence we have the input vector $\ket{b_1}=[0,1,0,0]^T=\ket{J^r_4}$. Similarly, the vectors $\ket{b_2}=[0,0,0,1]^T=\ket{J^r_1}+i\ket{J^r_2}$ and $\ket{b_3}=[1,0,0,0]^T=\ket{J^r_3}-i\ket{J^r_4}$ correspond to unit input signals from ports $\text{P}_2$ and $\text{P}_3$, respectively. By using Eq. (\ref{Eq:Greens1}), we can now obtain the Green's operator, and hence the linear response for each different excitation. For instance,  in the case of $\text{P}_1$, we obtain $G_4(\omega)\ket{f_1}=G_4(\omega)i\hat{\Gamma}\ket{b_1}=i\sqrt{2\gamma}\left[\frac{J^2k}{(\Delta\omega+i\gamma)^4}\ket{J^r_1}+\frac{Jk}{(\Delta\omega+i\gamma)^3}\ket{J_2^r}+\frac{J}{(\Delta\omega+i\gamma)^2}\ket{J_3^r}+\frac{1}{(\Delta\omega+i\gamma)}\ket{J_4^r}\right]$, where $\ket{f_1}=i\hat{\Gamma}\ket{b_1}$ and $\sqrt{2\gamma}$ corresponds to the element of the coupling matrix $\hat{\Gamma}$ between the input channels and the interior of the resonator system. Note that this expression describes the field amplitude inside the resonators system. To calculate the output, one must project this field amplitude on the output channel. For illustration purpose, let us consider the output channel $\text{Q}_1$, which is directly coupled to the $b_{CW}$, i.e. it corresponds to the vector $[0,0,0,1]^T=\ket{J_1^r}+i\ket{J_2^r}$. Thus one can assign the projection operator  $\mathbb{P}_1=\ket{J_1^r}\bra{\tilde{J}_1^l}+i\ket{J_2^r}\bra{\tilde{J}_2^l}$ to the output channel $\text{Q}_1$. Similarly, the projection operators $\mathbb{P}_2=\ket{J_4^r}\bra{\tilde{J}_4^l}$ and $\mathbb{P}_3=\ket{J_2^r}\bra{\tilde{J}_2^l}$ describe output channels $\text{Q}_2$ and $\text{Q}_3$, respectively. Finally, the output signal is given by: $\mathbb{P}_i\left[\hat{Y}\ket{b_j}-\hat{\Gamma}^T\hat{G}\ket{f_j}\right]$, where the matrix $\hat{Y}$ quantifies the direct coupling between the input and output channels whereas the matrix $\hat{\Gamma}^T$ describes the coupling between the interior of the resonators and the output channels. By applying the above-sketched recipe for an input/output from ports $P_1$ and $\text{Q}_3$, we find: $Q_3=-i\frac{2\gamma Jk}{(\Delta\omega+i\gamma)^3}$. In other words, counter-intuitively the response features a super-Lorentzian of order three, despite the fact that the system exhibits an EP of order four. This is, however, fully consistent with our theoretical analysis since $P_1$ corresponds to $\ket{J_4^r}$ and $Q_3$ corresponds to $\ket{J_2^r}$ as demonstrated by  mathematical structure of Eq. (\ref{Eq:Greens1}), and also pictorially in Fig. \ref{Fig:Dual_Space}\textbf{b}. Figure \ref{Fig:EP4_setup}\textbf{b} summarizes the steps descried in this section.\\

\noindent \textbf{DISCUSSION} \\
In this work, we have presented a detailed analysis of the linear response associated with non-Hermitian systems having EPs and showed that a non-diverging resolvent associated with the system's Hamiltonian can be expressed as an exact series expansion of the ordinary and generalized eigenfunctions of the Hamiltonian, i.e. without resorting to any perturbation approximation. Importantly, our formalism revealed a feature that escaped attention in previous studies, namely that the response lineshape scaling can be engineered by a judicious choice of the input and output channels. This observation is crucial for tailoring light-matter interactions at EPs. In order to emphasize this point and also clarify the application of our formalism, we have considered and analyzed several realistic photonic examples and we found excellent agreement between results obtained from full-wave simulations and our formulas. In doing so, we have also demonstrated an interesting effect analogous to adjoint coupling but rather in microring cavity setups. In other words, we have shown that more optical energy can be stored in a microring cavity system having an EP when the channel associated with the input signal matches the generalized eigenmode rather than the actual ordinary eigenmode of the structure. We emphasize that although the examples in this manuscript are chosen from the optical domain, our results are general and applicable to other physical systems that can be described by similar coupled mode formalism. These include electronic, acoustic, mechanical, and thermal systems. In addition, our framework provides a powerful tool for understanding the complex  interplay between non-Hermiticity and other physical effects such as topological invariants \cite{Hafezi_NatPhys2011, Schomerus2013OptLettTPM, Zhao2018THS, Resendiz2020PhysRevResearchTPN, Gong2018PhysRevX.8.031079, Feng_Science2019}, optomechanical coupling \cite{Jing2014SPL, Jing2015OIT, Jing2017HOEP, Zhang2018PLO, EisfledNJP2016}  
 as well as quantum statistics \cite{Schomerus2010PhysRevLettQNS, Arkhipov2019PhysRevA.99.053806, Nori_PRA2019, Metelmann2014PRL, Luitz2019PhysRevResearch.1.033051, Busch_PR2019}, to just mention a few examples. This in turn may enable the engineering of more elaborate schemes for controlling energy and information flow in complex non-Hermitian systems. Finally, we remark that understanding the linear response of non-Hermitian systems is a very crucial step towards studying their noise. In this regard, we expect our formalism to provide more insight into the noise behavior in non-Hermitian systems and play a positive role in the active debate on signal to noise ratio of EP-based sensors  \cite{LangbeinPhysRevA2018, LauNatComm2018, Zhang2019PhysRevLett.123.180501, Wang2020NatComm, PhysRevA.101.053846, David2020BeyondPetermann, Tsampikos2022EnhancedSignal}. We plan to explore some of these interesting directions in future works. 

{\section*{Data availability}
The data that support the findings of this study are available from the corresponding authors upon reasonable request.}



\begin{acknowledgments}
This work is supported by Air Force Office of Scientific Research (AFOSR) Multidisciplinary University Research Initiative (MURI) Award on Programmable systems with non-Hermitian quantum dynamics (Award No. FA9550-21-1-0202). R.E. also acknowledges support from ARO (Grant No. W911NF-17-1-0481), NSF (Grant No. ECCS 1807552), and the Alexander von Humboldt Foundation. S.K.O. acknowledges support from NSF (Grant No. ECCS 1807485)
K.B. acknowledges support by the Leibniz Association within the SAW-project LAPTON and from the Alexander von Humboldt Foundation.
\end{acknowledgments}

\section*{Author contribution}
R.E. conceived the project. R.E. and A.H. performed the theoretical analysis with feedback from S.K.O, K.B., and D.N.C. The numerical simulations was performed by A.H. All authors contributed to the manuscript writing.

\section*{Competing interests}
The authors declare no competing interests

\newpage

\begin{center}
\textbf{{\Large Supplementary Information:}}

\textbf{{\Large Linear response theory of open systems with exceptional points}}

{A. Hashemi, K. Busch, D. N. Christodoulides, S.K. Ozdemir, and R. El-Ganainy}
\end{center}

\newpage

\section*{SUPPLEMENTARY NOTES}
\subsection{Subtleties arising in resonant systems with EPs: an example}
Here we provide a detailed analysis for the example shown in Fig.\:2 in the main text, which is also shown here for convenience. It consists of three identical microring resonators that are coupled sequentially via horizontal waveguides. An additional vertical waveguide provides access to selectively excite the second resonator. We neglect the cross talk between the horizontal and vertical waveguides since it can be minimized using various design strategies \cite{Daly1996JLT,Mingaleev2004OptLett,Kobayashi2005OR,Longhi2015OptLett}. A similar system was considered in \cite{Qi2020PRApplied} and shown to exhibit a third order EP in the subspace spanned by the CW, CCW and CW modes of the resonators $\text{R}_{1,2,3}$, respectively. In this subspace, and by allowing excitations only from ports $P_{1,2,3}$ with the collection output ports $Q_1$ and $Q_2$ (see supplementary Fig. \ref{Fig_Three_Microring}) the system is described by the coupled equations:
\begin{align}\label{Eq:three_microring}
i\frac{d}{dt}
\left[\begin{matrix}
a_1\\
a_2\\
a_3
\end{matrix}\right]
=&\left[\begin{matrix}
\omega_o-3i\gamma & 0 & 0\\
\kappa & \omega_o-3i\gamma & 0\\
0 & \kappa & \omega_o-3i\gamma
\end{matrix}\right]
\left[\begin{matrix}
a_1\\
a_2\\
a_3
\end{matrix}\right]
+i\sqrt{2\gamma}\left[\begin{matrix}
P_1\\
P_2\\
P_3
\end{matrix}\right],\nonumber\\
Q_1=&P_3-\sqrt{2\gamma}\:a_3,\nonumber\\
Q_2=&P_I-\sqrt{2\gamma}\:a_3,
\end{align}
where $a_{1,2,3}$ are modal amplitudes of the CW, CCW and CW modes associated with resonators $\text{R}_{1,2,3}$, respectively. The resonant frequency and decay rate of each resonator are given by  $\omega_o$ and $3 \gamma$. For resonator $\text{R}_2$ the decay rate is due to equal coupling to three waveguides. On the-other hand, for $\text{R}_{1,3}$ the coupling to waveguides contributes only $2\gamma$ and we assume that an additional loss of $\gamma$ is intentionally introduced for example by depositing a metal layer on top of the ring or introducing an auxiliary waveguide next to each existing waveguide. The ports $P_{1,2,3}$ serve as excitation channels for the modes $a_{1,2,3}$, respectively, while the output signal is collected from ports $Q_1$ and $Q_2$ as depicted in supplementary Fig. \ref{Fig_Three_Microring}. In addition, $P_I$ represents the signal in the waveguide between $\text{R}_2$ and $\text{R}_3$ (see supplementary Fig. \ref{Fig_Three_Microring}\textbf{b}). Within this modal subspace, there is unidirectional coupling from modes $a_1\rightarrow a_2$ and $a_2\rightarrow a_3$. The exact details of this waveguide-mediated indirect coupling depends on the evanescent tunneling between the rings and waveguides as well as the distances between the rings. Here we assume that unidirectional coupling coefficients are identical and denote them by $\kappa$.

\begin{figure}
	\includegraphics[width=\textwidth]{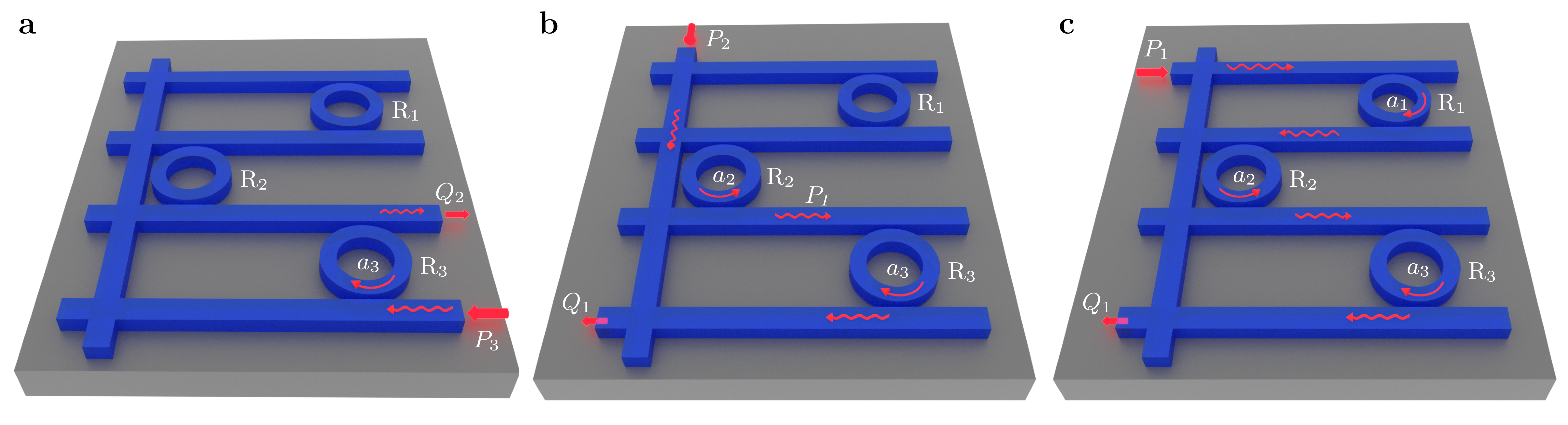}
	\caption{\textbf{Subtleties of linear response of non-Hermitian systems having EPs: an illustrative example.} A photonic system that exhibits an EP of order three. As described in the main text, it can have very different linear responses based on the input/output channel configuration. Here we present the detailed mathematical analysis of this structure.}
	\label{Fig_Three_Microring}
\end{figure}

In this case, the resolvent, $G_3$, can be evaluated in closed form:
\begin{equation}
G_3=\left[
\begin{array}{ccc}
\vspace*{2mm}
 \mu_3 & 0 & 0 \\
 \vspace*{2.5mm}
 \kappa \mu_3^2 & \mu_3 & 0 \\
 \kappa ^2\mu_3^3 &\kappa \mu_3^2 & \mu_3 \\
\end{array}
\right],
\end{equation}
\noindent where $\mu_3\equiv 1/(\omega-\omega_o+3i\gamma)$. One can easily confirm that the exceptional eigenvector of $G_3$ is $\ket{K_1^r}=[0,0,1]^T$, while the generalized eigenvectors forming the Jordan chain are give by $\ket{K_2^r}=[0,\frac{\xi}{\kappa\mu_3^2},0]^T$ and $\ket{K_3^r}=\left[(\frac{\xi}{\kappa\mu_3^2})^2,-\frac{\chi _o^2}{\kappa \mu_3^3} ,0\right]^T$ (see Eq.(5.1) of supplementary note 6). \\

Next, we consider the above system when it is excited from one of the ports $P_{1,2,3}$. These signal pathways inside the structure for each of these distinct situations are depicted in supplementary Figs. \ref{Fig_Three_Microring}\textbf{a}, \textbf{b}, and \textbf{c}. It is straightforward to show that the scattering coefficient between the input and output ports in each case is given by:
\begin{align}
\begin{split}
\frac{Q_2}{P_3}&=-2i\gamma \mu_3=-\frac{2i\gamma}{(\omega-\omega_o+3i\gamma)},\\
\frac{Q_1}{P_2}&=-2i\gamma \kappa \mu_3^2=-\frac{2i\gamma \kappa}{(\omega-\omega_o+3i\gamma)^2},\\
\frac{Q_1}{P_1}&=-2i\gamma \kappa^2 \mu_3^3=-\frac{2i\gamma \kappa^2}{(\omega-\omega_o+3i\gamma)^3}.
\end{split}
\end{align}
These results can be intuitively understood by referring again to supplementary Fig. \ref{Fig_Three_Microring}, which depicts the designated input and output ports, and the actual paths taken by the excitation signal. In the first case, the input signal $P_3$ crosses only the ring resonator $\text{R}_3$ before it couples to the output $Q_2$. On the other hand, the signal $P_2$ in the second case interacts with two rings ($\text{R}_2$ and $\text{R}_3$) before it is coupled to the output channel $Q_1$. Finally, the trajectory of the input signal $P_1$ in the third case involves three resonators and the output port $Q_1$. The results discussed above can be understood by noting that each ring contributes a Lorentzian response and that total response is the product of the individual responses (due to the series connection of the rings). In more complex systems where such a simple trajectory picture does not exist, it is necessary to develop a general theoretical framework to analyze and understand the system's response.

\subsection{Completeness and biorthogonality of the eigenvectors associated with defective matrices }\label{Appendix:bi-orthogonality}
The eigenvectors of non-Hermitian matrices are in general non-orthogonal. Instead, they obey a modified bi-orthogonality relation, i.e. orthogonality between the right eigenvector and its corresponding left eigenvectors. The spectral analysis of a non-Hermitian matrix becomes rather complicated when it is defective, i.e. exhibiting EPs in its spectrum. In what follows, we provide a concise discussion that elaborates on this situation. Let us assume an $N\times N$ matrix $\hat{H}_{EP}$ that has one EP of order $M$ (generalization to multiple EPs is straightforward). The right eigenvectors ($\ket{\psi_n^r}$) of $\hat{H}_{EP}$ are defined by

\begin{equation} \label{Eq:Right_Evectors}
(\hat{H}_{EP}-\Omega_n\hat{I})\ket{\psi_n^r}=0,
\end{equation}

\noindent where $n=1,2,\ldots,N-M$. Evidently, these vectors do not span the underlying vector space. It is well-known however that the bases can be completed by using the Jordan canonical chain procedure, i.e. by defining the generalized right eigenvectors ($\ket{J^r_m}$) of $\hat{H}_{EP}$ by the Jordan chain

\begin{equation} \label{Eq:Right_G_Evectors}
(\hat{H}_{EP}-\Omega_{EP}\hat{I})\ket{J_m^r}=\chi_m\ket{J_{m-1}^r},
\end{equation}

\noindent where $m=1,2,\ldots, M$ and $\chi_1=0$. The eigenvectors $\ket{J_m^r}$ are linearly independent. This can be shown as follows. Consider the relation $\sum_{m=1}^{M} z_m \ket{J_m^r}=0$. We now apply the operator $(H-\Omega_{EP}\hat{I})^{M-1}$ to both sides. All the terms on the left-hand-side will vanish except the last term, which gives $z_M=0$. By repeating this procedure on the remaining terms, we find that all the coefficients $z_m$ have to be zero. Similarly, one can show that any vector $\ket{J_m^r}$ is linearly independent from all the vectors $\ket{\psi_{n}^r}$. In other words, the set $\{\ket{\psi_n^r}\}$ spans an $N-M$ dimensional space (call it $\mathcal{D}_{\phi}$) while the set $\{\ket{J_n^r}\}$ spans the complementary $M$ dimensional space $\mathcal{D}_{EP}$. Taken together, the set $\{\ket{\psi_n^r}, \ket{J_n^r}\}$ forms a complete basis that spans $\mathcal{D}_{\phi} \cup \mathcal{D}_{EP}$. Thus any vector in that space can be decomposed in terms of these bases. However, as we mentioned, in general $\braket{\psi_m^r|\psi_n^r}\neq \delta_{m,n}$. This, in turn complicates the procedure for finding the projection of any general vector onto the bases vectors. This difficulty can be overcome by using the concept of left eigenvectors:

\begin{equation} \label{Eq:Left_Evectors}
\bra{\psi_n^l}(\hat{H}_{EP}-\Omega_n\hat{I})=0.
\end{equation}

\noindent The set $\{\bra{\psi_m^l}\}$ defines a dual space of $\mathcal{D}_{\phi}$ (see Fig.\:5\textbf{a}). From Eq.\:(\ref{Eq:Right_Evectors}), we can write $\braket{\psi_{n_2}^l|\hat{H}_{EP}|\psi_{n_1}^r}=\Omega_{n_1}\braket{\psi_{n_2}^l|\psi_{n_1}^r}$. Similarly, from Eq.\:(\ref{Eq:Left_Evectors}) we obtain $\braket{\psi_{n_2}^l|\hat{H}_{EP}|\psi_{n_1}^r}=\Omega_{n_2}\braket{\psi_{n_2}^l|\psi_{n_1}^r}$. For these two relations to be consistent for $\Omega_{n_1}\neq \Omega_{n_2}$, we must have $\braket{\psi_{n_2}^l|\psi_{n_1}^r}=0$ for $n_1\neq n_2$. This is known as biorthogonality. By using this last relation iteratively in the recursive system of Eqs.\:(\ref{Eq:Right_G_Evectors}), we also find that $\braket{\psi_m^l|J_n^r}=0$ for any $m$ and $n$.  For instance, by using $(\hat{H}_{EP}-\Omega_{EP}\hat{I})^2\ket{J_2^r}=0$, we find that $\braket{\psi_m^l|(\hat{H}_{EP}-\Omega_{EP}\hat{I})^2|J_2^r}=0$. On the other hand, by using $\bra{\psi_m^l}(\hat{H}_{EP}-\Omega_{EP}\hat{I})^2=\bra{\psi_m^l}(\Omega_m-\Omega_{EP})^2$, the same expression can be evaluated to be $\braket{\psi_m^l|(\hat{H}_{EP}-\Omega_{EP}\hat{I})^2|J_2^r}=(\Omega_m-\Omega_{EP})^2 \braket{\psi_m^l|J_2^r}$, which gives $\braket{\psi_m^l|J_2^r}=0$.\\

Similarly, one can define the left generalized eigenvectors of $\hat{H}_{EP}$ by following the same procedure, namely:

\begin{equation} \label{Eq:Left_G_Evectors1}
\bra{J_m^l}(\hat{H}_{EP}-\Omega_{EP}\hat{I})={\chi}_{m}\bra{J^l_{m-1}}.
\end{equation}

\noindent Following the same procedure as before, one can show that the vectors $\bra{J_m^l}$ are linearly independent from each other and from $\bra{\psi_n^l}$, i.e. they define a dual space of $\mathcal{D}_{EP}$ (see Fig.\:5\textbf{a} in the main text). Also, it is straightforward to show that $\braket{J_m^l|\psi_n^r}=0$ for any $m$ and $n$. On the other hand, the relation between the vectors $\bra{J^l}$ and $\ket{J^r}$ is subtle due to the self-orthogonality:  $\braket{J_m^l|J^r_m}=0$ for $m\leq \frac{M}{2}$. This can be proven by considering the expression $\braket{J^l_m|(\hat{H}_{EP}-\Omega_{EP}\hat{I})^m|J^r_{2m}}$. When evaluated by first calculating the term $\bra{J^l_m}(\hat{H}_{EP}-\Omega_{EP}\hat{I})^m$, we find that the result is zero. On the other hand, when we first evaluate the term $(\hat{H}_{EP}-\Omega_{EP}\hat{I})^m \ket{J^r_{2m}}$, we obtain $\chi_{2m}\chi_{2m-1}\ldots\chi_{m+1}\braket{J^l_m|J^r_m}$, which proves the self-orthogonality relation. \\

In order to overcome the above problem, we first note that the set of the left eigenvectors $\{\bra{J_m^l}\}$ spans the dual space of $\mathcal{D}_{EP}$. Thus any vector formed by an arbitrary superposition of  $\{\bra{J_m^l}\}$ also lies in the dual space of $\mathcal{D}_{EP}$. Let us now define a new set of vectors $\bra{\tilde{J}_m^l}$, each of which lies in the dual space of $\mathcal{D}_{EP}$, i.e. $\bra{\tilde{J}_m^l}=\sum_{k=1}^{M}t_{m,k}\bra{J_k^l}$. In order to find the $M^2$ coefficients $t_{m,k}$, we need to impose $M^2$ normalization conditions. These can be chosen to satisfy $\braket{\tilde{J}_m^l|J_k^r}=\delta_{m,k}$. Note that the new vectors $\bra{\tilde{J}_m^l}$ are not generalized eigenvectors of the Hamiltonian $\hat{H}_{EP}$.

\subsection{Simultaneous normalization of right and left eigenvectors}\label{Appendix:normal matrix}
The simultaneous normalization of the right and left eigenvectors associated with a non-Hermitian Hamiltonian was discussed in \cite{Siegman1989PRA}. For completeness, we reiterate some of these results here from a more general point of view. The main result of this section is that for non-normal matrices (matrices that do not commute with their Hermitian conjugates), the three different normalization conditions $\braket{\psi_n^l|\psi_n^r}=1$, $\braket{\psi^r_n|\psi^r_n}=1$, and $\braket{\psi_n^l|\psi_n^l}=1$ cannot be satisfied simultaneously. For non-defective Hamiltonians, this is a direct consequence of Cauchy-Schwarz inequality: $|\braket{\psi_n^l|\psi_n^r}|^2\leq \braket{\psi_n^r|\psi_n^r}\braket{\psi_n^l|\psi_n^l}$. The equality holds if and only if the vectors $\ket{\psi_n^r}$ and $\ket{\psi_n^l}$ are the same up to a constant. This occurs when $\hat{H}$ is normal, i.e. $[\hat{H},\hat{H}^{\dagger}]=0$. To prove this, we note that the above relation implies that both $\hat{H}$ and $\hat{H}^{\dagger}$ share the same right and left eigenvectors. It follows that $\ket{\psi_n^r}$ and $\bra{\psi_n^l}$ are also right and left eigenvectors of the operator $\hat{H}\hat{H}^{\dagger}$. This last operator is Hermitian and thus, up to a constant, the right and left eigenvectors are identical. \\

\subsection{Series expansion of the resolvent}\label{Appendix:response derivation}
In this section, we present a detailed derivation for the series expansion of the resolvent in Eq.\:(6). By considering an arbitrary input signal $\ket{u}=\sum_{n=1}^{N-M} c_n \ket{\psi_n^r}+\sum_{m=1}^{M} d_m \ket{J_m^r}$, where the coefficients $c_n$ and $d_m$ are known (or can be calculated using the projection rules for the right/left eigenvectors as discussed in the previous section), and by assuming a similar series representation of the response signal with known coefficients $\tilde{c}_n$ and $\tilde{d}_m$, we can write:

\begin{align}\label{Eq:Fourier sol}
(\omega \hat{I}-\hat{H}_{EP})\left(\sum_{n=1}^{N-M} \tilde{c}_n\ket{\psi_n^r}+\sum_{m=1}^{M} \tilde{d}_m\ket{J_m^r}\right)=
\sum_{n=1}^{N-M} c_n \ket{\psi_n^r}+\sum_{m=1}^{M} d_m \ket{J_m^r}.
\end{align}

\noindent By using Eqs.\:(\ref{Eq:Right_Evectors}) and (\ref{Eq:Right_G_Evectors}), we can express the left-hand-side of Eq.\:(\ref{Eq:Fourier sol}) as $\sum_{n=1}^{N-M} \tilde{c}_n(\omega-\Omega_n)\ket{\psi_n^r}+\sum_{m=1}^{M} \left[\tilde{d}_m(\omega-\Omega_{EP})-\tilde{d}_{m+1}\chi_{m+1}\right]\ket{J_m^r}$. By recalling that the vectors $\ket{\psi_n^r}$ and $\ket{J_m^r}$ are all linearly independent, we arrive at:

\begin{align}
\tilde{c}_n&=\frac{c_n}{\omega-\Omega_n}\label{Eq:c coefficient}\\
\tilde{d}_m&=\frac{d_m}{\omega-\Omega_{EP}}+\frac{\chi_{m+1}\tilde{d}_{m+1}}{\omega-\Omega_{EP}}.\label{Eq:recursive}
\end{align}

\noindent The recursive relation in Eq.\:(\ref{Eq:recursive}) can be further simplified by noting that $\chi_{M+1}=0$, which gives $\tilde{d}_M=\frac{d_M}{\omega-\Omega_{EP}}$. By inserting this result back in Eq.\:(\ref{Eq:recursive}), we obtain the general expression:

\begin{align}\label{Eq:d coefficient}
\tilde{d}_m =&
\frac{d_m}{\omega-\Omega_{EP}} + \frac{\chi_{m+1}d_{m+1}}{(\omega-\Omega_{EP})^2}+
\frac{\chi_{m+1}\chi_{m+2}d_{m+2}}{(\omega-\Omega_{EP})^3}+\cdots+ \frac{\chi_{m+1}\chi_{m+2}\cdots\chi_{M}d_M}{(\omega-\Omega_{EP})^{M-m+1}}\nonumber\\
=&\sum_{k=m}^{M}\alpha_k^{(m)}\frac{d_k}{(\omega-\Omega_{EP})^{k-m+1}},
\end{align}

\noindent where we defined $\alpha_m^{(m)}=1$ and $\alpha_k^{(m)}=\alpha_{k-1}^{(m)}\chi_k=\chi_k\chi_{k-1}\ldots\chi_{m+1}$ for $k=m+1,\ldots,M$. Finally, by using $c_n=\braket{\psi_n^l|u},\ d_k=\braket{\tilde{J}_k^l|u}$ (see supplementary note \ref{Appendix:bi-orthogonality}) we obtain:

\begin{align}
\hat{G}_{EP}\ket{u}&=\biggr(\sum_{n=1}^{N-M}\frac{\ket{\psi^r_n}\bra{\psi_n^l}}{\omega-\Omega_n}+
\sum_{m=1}^{M}\sum_{k=m}^{M}\alpha_k^{(m)}\frac{\ket{J^r_m}\bra{\tilde{J}^l_k}}{(\omega-\Omega_{EP})^{k-m+1}}\biggr)\ket{u}.
\end{align}

\noindent Since $\ket{u}$ is an arbitrary vector, it follows that the series summation on the right-hand-side represents the resolvent which proves Eq.\:(6).

\subsection{Non-interfering excitation channels}
Here we discuss the classification of the excitation channels based on their interference properties. Consider an input $\ket{u}= \ket{u_1}+ \ket{u_2}$, with $s(t)=e^{-i\omega_e t}$. This rather unphysical assumption of an input signal that extends over the time axis is well suited for treating long time limits after the transient response has died out, which is the case of interest here (alternatively, one can, of course, use the Laplace transform and take the long-time limit $t \rightarrow \infty$ explicitly). The normalized energy (with respect to some reference energy value), stored in the system after the transient response fades away, is given by $\braket{E}\equiv\lim_{t\rightarrow \infty} \overline{\braket{a(t)|a(t)}}$,  with the overline indicating time average. Equivalently, the normalized stored energy can be also expressed as $\braket{E}= \braket{\tilde{A}(\omega_e)|\tilde{A}(\omega_e)}$ where $\ket{A(\omega)}=\ket{\tilde{A}(\omega_e)}\delta(\omega-\omega_e)$, or $ \ket{\tilde{A}(\omega_e)}=\sum_{n=1}^{N} \frac{c_n}{\omega_e-\Omega_n}\ket{\psi_n^r}$.  In general, the above expression will contain contributions from (1) $\ket{u_1}$ only in the absence of $\ket{u_2}$; (2) $\ket{u_2}$ only in the absence of $\ket{u_1}$; (3) interference component due to the non-orthogonality of $\ket{u_{1,2}}$. To illustrate this with a concrete example, consider an input $\ket{f}=\left[\alpha_m \ket{\psi_m^r}+\alpha_n \ket{\psi_n^r}\right] e^{-i\omega_e t}$. The normalized stored energy in this case is given by $\braket{E}=|\frac{\alpha_m}{\omega_e-\Omega_m}|^2 \braket{\psi_m^r|\psi_m^r}+|\frac{\alpha_n}{\omega_e-\Omega_n}|^2 \braket{\psi_n^r|\psi_n^r}+2 \text{Re}\{\frac{\alpha_m^*}{\omega_e-\Omega_m^*} \frac{\alpha_n}{\omega_e-\Omega_n} \braket{\psi_m^r|\psi_n^r}\})$, with the last term representing the interference term. This raises the question of whether this feature is pertinent to any input profile or if it is possible to construct some excitation channels whose energy contribution inside the resonators do not interfere. The main result of this section indeed affirms the latter possibility. To demonstrate this, we first note that the resolvent $\hat{G}_{EP}(\omega)$ has the same spectral structure of the Hamiltonian $\hat{H}_{EP}$, i.e. they have the same eigenvectors and their eigenvalues are given by $\frac{1}{\omega-\Omega_n}$ and $\Omega_n$ respectively. Consequently, they also share the same EP (see supplementary note 5). As a result, one can in principle complete the basis by constructing the Jordan vectors associated with $\hat{G}_{EP}(\omega)$ instead of $\hat{H}_{EP}$ as we have done before:
\begin{equation} \label{Eq:JordanChain_G}
\begin{aligned}
(\hat{G}_{EP}-\mu_{EP}\hat{I})\ket{K_1^r} &=0  \\
(\hat{G}_{EP}-\mu_{EP}\hat{I})\ket{K_2^r} &=\xi\ket{K_1^r} \\
&\vdots \\
(\hat{G}_{EP}-\mu_{EP}\hat{I})\ket{K_M^r} &=\xi\ket{K_{M-1}^r}.
\end{aligned}
\end{equation}
\noindent In the above, $\mu_{EP}=\frac{1}{\omega-\Omega_{EP}}$ and the coefficient $\xi$ is taken to be of unit value and of dimensions similar to $\mu_{EP}$. We emphasize that both $\hat{G}_{EP}$ and $\mu_{EP}$ are functions of $\omega$. It is straightforward to show that $\ket{K_1^r}=\ket{J_1^r}$. A more general Jordan vector $\ket{K_m^r}$ with $m>1$, can be expressed as a linear superposition of the vectors $\ket{J^r_m}$ (see supplementary note 6). We now consider the case when the matrix representation of $\hat{G}_{EP}$ is given by the Jordan canonical form. In this case, the generalized eigenvectors are orthogonal. This is of course a feature of the geometry of the problem. For instance, if the effective Hamiltonian of the optical structure under study is given by $\hat{H}=$\resizebox{0.6in}{!}{$
\begin{pmatrix}
	\Omega & J & \kappa\\
	0 & \Omega & J\\
	0 & 0 & \Omega
	\end{pmatrix}$}, then we find $\hat{G}(\omega=\Omega-\frac{J^2}{\kappa})=\frac{\kappa^2}{J^3}$\resizebox{0.8in}{!}{$\begin{pmatrix}
-\frac{J}{\kappa} & 1 & 0\\
0 & -\frac{J}{\kappa} & 1\\
0 & 0 & -\frac{J}{\kappa}
\end{pmatrix}$} with $\Ket{K^r_1}=[1, 0, 0]^T$, $\Ket{K^r_2}=[0, 1, 0]^T$, and $\Ket{K^r_3}=[0, 0, 1]^T$.
By assuming that the above condition is satisfied, we now consider an input of the form $\ket{f}=\left[\alpha_m \ket{K_m^r}+ \alpha_n \ket{K_n^r}\right] e^{-i\omega_e t}$. Clearly, if $|n-m|=1$, interference terms will appear in the expression for $\braket{E}$. On the other hand, when $|n-m|>1$, the energy expression will not contain any interference terms. For instance, for $m=2$ and $n=4$, we will obtain $\braket{E}=\braket{E_2}+\braket{E_4}$, where $\braket{E_2}=|\alpha_2|^2\left[|\xi|^2\braket{K_1^r|K_1^r}+|\mu_{EP}(\omega_e)|^2 \braket{K_2^r|K_2^r}\right]$ and $\braket{E_4}=|\alpha_4|^2\left[|\xi|^2\braket{K_3^r|K_3^r}+|\mu_{EP}(\omega_e)|^2 \braket{K_4^r|K_4^r}\right]$. If, on the other hand, the natural matrix representation (i.e. the representation that arises from the geometry of the problem without performing any linear mapping) of $\hat{G}$ is not in the Jordan canonical form, the generalized eigenvectors do not have to be orthogonal. However, by using the similarity transformation that relates the natural and canonical bases one can find the non-interfering channels in the natural bases. These in general will be a superposition between the excitation channels associated with the natural bases.\\
\indent The above analysis in terms of the Jordan vectors associated with the resolvent illustrates another interesting effect. Consider the two different inputs  $\ket{f_{1,2}}=\alpha_{1,2}\ket{K_{1,2}^r} e^{-i\omega_e t}$, with the constants $\alpha_{1,2}$ chosen to satisfy the conditions $|\alpha_{1,2}|^2\braket{K_{1,2}^r|K_{1,2}^r}=1$. In this case, the expressions for the corresponding energy stored in the structure are given by: $\braket{E_1}=|\mu_{EP}(\omega_e)|^2$, and $\braket{E_2}=|\mu_{EP}(\omega_e))|^2 +|\xi|^2|\alpha_2/\alpha_1|^2$. In other words, $\braket{E_2}/\braket{E_1}=1+\frac{\xi^2}{|\mu_{EP}|^2} |\alpha_2/\alpha_1|^2>1$. This result indicates that in this case, mode matching does not lead to the most efficient excitation scheme- a feature that resembles the notion of adjoint coupling in the context of free space unstable laser resonators \cite{Siegman1995APB}.

\subsection{Generalized eigenvectors of the resolvent}\label{Appendix:Green's operator eigenvectors}
A direct consequence of resolvent definition is that it shares the same eigenvectors with its corresponding Hamiltonian, including the exceptional vector, i.e. $\ket{K_1^r}=\ket{J_1^r}$. However, the generalized eigenvectors of the Hamiltonian and those of the corresponding resolvent are not identical. Additionally, as we have mentioned before, $\ket{K_n^r}$ and $\ket{J_n^r}$ are not unique. However, once the two sets of vectors are chosen according to a specific criterion, one can express the former in terms of the latter. In particular, one can write $\ket{K_m^r}=\sum_{i=1}^{M} C_i^m \ket{J_i^r}$, where the coefficients $C_i^m$ are given by $C_i^m=\braket{\tilde{J}_i^l|K_m^r}$ (see supplementary note \ref{Appendix:bi-orthogonality} for the discussion of the left vectors $\bra{\tilde{J}_i^l}$). \\
\indent In general, the above series expansion should be computed numerically. However, one particular solution can be expressed in closed form. In the following, we derive this result. Consider a choice of the vectors $\ket{K_m^r}$ as $\ket{K^r_m}=\sum_{i=2}^m C_i^m\ket{J^r_i}$, with $m=2,3,\ldots,M$. Note that the upper limit of the summation is $m$, not $M$, i.e. it varies from one vector to another. The strategy is to substitute this ansatz into the formula $(\hat{G}_{EP}-\mu_{EP}\hat{I})\ket{K^r_m}=\xi\ket{K^r_{m-1}}$ and solve for the vectors $\ket{K_m^r}$. Before we proceed, we recall $\mu_{EP}\equiv(\omega-\Omega_{EP})^{-1}$. Let us now consider the case $m=2$. It follows that $(\hat{G}_{EP}-\mu_{EP}\hat{I})C_2^2\ket{J^r_2}=\xi\ket{J^r_{1}}$. By multiplying both sides by $1/\hat{G}_{EP} \equiv (\omega\hat{I}-\hat{H}_{EP})$, we obtain $\mu_{EP}C_2^2\chi_2\ket{J_1^r}=\xi(\omega-\Omega_{EP})\ket{J_1^r}$, which in turn leads to:

\begin{equation} \label{C22}
C_2^2=\frac{\xi}{\mu_{EP}^2\chi_2}.
\end{equation}

\noindent By following similar steps for $m>2$, we arrive at the relation $\sum_{i=2}^m C_i^m\mu_{EP}\chi_i\ket{J^r_{i-1}}=\xi\sum_{i=2}^{m-1} \frac{C_i^{m-1}}{\mu_{EP}}\ket{J^r_i}-\xi\sum_{i=2}^{m-1} C_i^{m-1}\chi_i\ket{J^r_{i-1}}$. By changing the summation index $i \rightarrow i+1$ in the first and third terms, the above formula reads $\sum_{i=1}^{m-1} C_{i+1}^m\mu_{EP}\chi_{i+1}\ket{J^r_{i}}=\xi
\sum_{i=2}^{m-1} \frac{C_i^{m-1}}{\mu_{EP}}\ket{J^r_i}-\xi\sum_{i=1}^{m-2} C_{i+1}^{m-1}\chi_{i+1}\ket{J^r_{i}}$. By rearranging the terms, we obtain $\left(C_{2}^m\mu_{EP}\chi_{2}+\xi C_{2}^{m-1}\chi_{2}\right)\ket{J^r_{1}}+\sum_{i=2}^{m-2} \left(C_{i+1}^m\mu_{EP}\chi_{i+1}- \xi\frac{C_i^{m-1}}{\mu_{EP}}+\xi C_{i+1}^{m-1}\chi_{i+1} \right)\ket{J^r_{i}}+ \left(C_m^m\mu_{EP}\chi_m-\xi\frac{C_{m-1}^{m-1}}{\mu_{EP}} \right)\ket{J^r_{m-1}}=0$. By invoking the linear independence of vectors $\ket{J_i^r}$ and using the above expression for $C_2^2$ , we arrive at the following expressions for the coefficients $C_2^m$ and $C_m^m$:

\begin{align} \label{Eq:CMM}
C_2^m &=\left(-\frac{1}{\mu_{EP}}\right)^m\frac{\xi^{m-1}}{\chi_2}\nonumber\\
C_m^m &=\left(\frac{\xi}{\mu_{EP}^2}\right)^{m-1}\frac{1}{\chi_2\chi_3\ldots\chi_m}
\end{align}

together with the recursive relation:

\begin{align} \label{Eq:Recursive}
C^m_i=\left(\frac{1}{\mu_{EP}}\right)^2\frac{\xi}{\chi_i}C_{i-1}^{m-1}-\frac{\xi}{\mu_{EP}}C_i^{m-1},
\end{align}

\noindent where $i=3,4,\ldots,m-1$. Relation (\ref{Eq:Recursive}) is a recursive equation with variable coefficients. The unknown expansion constants $C_i^m$ in Eq.\:(\ref{Eq:Recursive}) can be obtained by the aid of the known coefficients in Eq.\:(\ref{Eq:CMM}). For instance, the value of $C_3^4$ depends on those of $C_2^3$ and $C_3^3$ which are obtained from the first and second lines of Eq.\:(\ref{Eq:CMM}). In general, one can confirm by direct substitution that the solution of Eq.\:(\ref{Eq:Recursive}) can be written as:

\begin{align}
C_i^m=(-1)^{m+i}
\begin{pmatrix}
m-2\\
i-2
\end{pmatrix}
\frac{\mu_{EP}^{2-m-i}}{\chi_2\chi_3\ldots\chi_i}\xi^{m-1},
\end{align}

\noindent where $\begin{pmatrix}m\\n\end{pmatrix}\equiv\frac{m!}{n!(m-n)!}$ is the binomial coefficient. Thus we finally arrive at:

\begin{align}
\ket{K^r_m}=\sum_{i=2}^m (-1)^{m+i}
\begin{pmatrix}
m-2\\
i-2
\end{pmatrix}
\frac{\mu_{EP}^{2-m-i}\xi^{m-1}}{\chi_2\chi_3\ldots\chi_i} \ket{J^r_i}.
\end{align}

\noindent It is worth noting that the expansion coefficients obtained above are functions of the frequency $\omega$ of the input drive signal.

\subsection{Systems with multiple exceptional points}\label{Appendix:multiple_EP}
Let us assume a Hamiltonian with two exceptional points of order, say, $M_1$ and $M_2$ and energy eigenvalues $\Omega_{EP}^{(1)}$, $\Omega_{EP}^{(2)}$ respectively (the case with several exceptional points is a straightforward generalization). The generalized eigenvalue equations for the left and right generalized eigenvectors associated with the two EPs are given by:

\begin{align}
	\left(\hat{H}_{EP}-\Omega_{EP}^{(1)}\right)\ket{J_{m}^{r}} &=\chi_{m}^{(1)}\ket{J_{m-1}^{r}}\label{Eq:right1}\\
	\bra{J_{m}^l}\left(\hat{H}_{EP}-\Omega_{EP}^{(1)}\right) &=\chi_{m}^{(1)}\bra{J_{m-1}^{l}}\label{Eq:left1}
\end{align}

and

\begin{align}
	\left(\hat{H}_{EP}-\Omega_{EP}^{(2)}\right)\ket{I_{k}^{r}} &=\chi_{k}^{(2)}\ket{I_{k-1}^{r}}\label{Eq:right2}\\
	\bra{I_{k}^l}\left(\hat{H}_{EP}-\Omega_{EP}^{(2)}\right) &=\chi_{k}^{(2)}\bra{I_{k-1}^{l}},\label{Eq:left2}
\end{align}

\noindent where $m=1,2,\ldots,M_1$ and $k=1,2,\ldots,M_2$ and $\chi_1^{(1,2)}=0$. In what follows, we show that the generalized eigenvectors associated with the two different EPs are biorthogonal. Let us start by inspecting the expression $\braket{J_m^l|\left(\hat{H}_{EP}-\Omega_{EP}^{(1)}\right)^m|I_1^r}$. It can be evaluated by first applying the operator $\left(\hat{H}_{EP}-\Omega_{EP}^{(1)}\right)^m$ to the vector $\ket{I_1^r}$ from the left or by acting on the vector $\bra{J_m^l}$ from the right. By doing so,  we obtain $\braket{J_m^l|I_1^r}=0$. Similarly, one can show that $\braket{J_1^l|I_k^r}=0$.\\

\begin{figure}
	\includegraphics[scale=0.9]{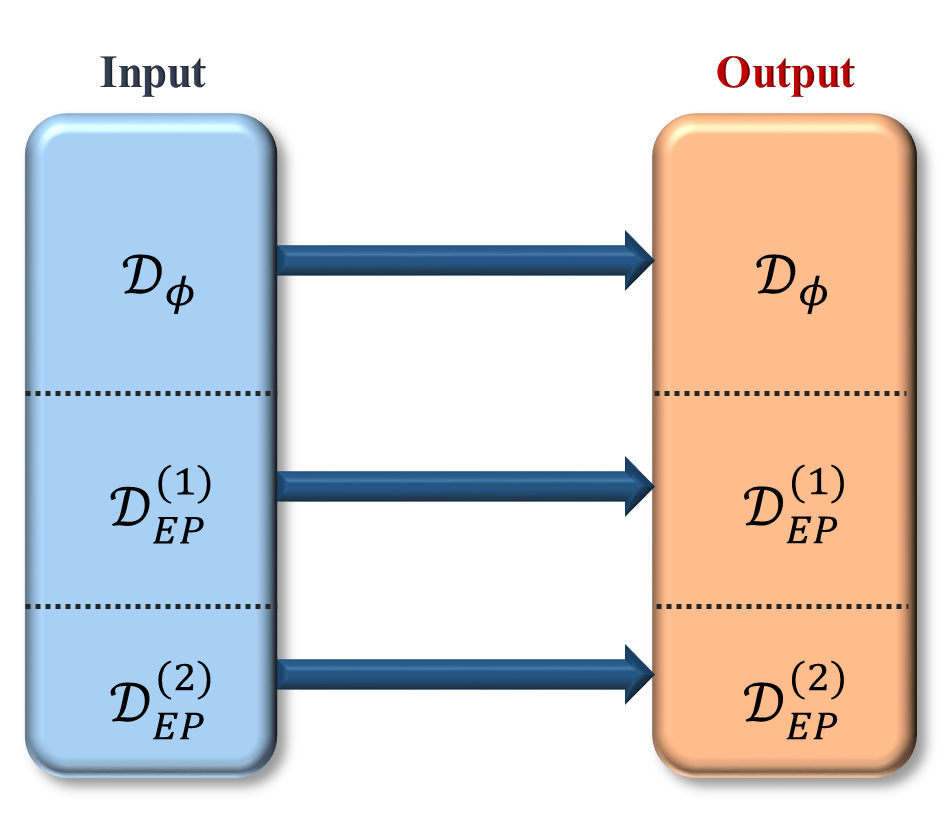}
	\caption{General structure of the linear response associated with a non-Hermitian system that exhibits two different EPs. As can be seen, each domain is coupled only to itself. In other words, there is no coupling between the two sub-spaces spanned by the generalized eigenvectors associated with the two EPs. Thick arrows indicate collective coupling between the domains. The details of the coupling (not shown here) follow a scheme identical to that depicted in Fig.\:5\textbf{b}.}
	\label{Fig:Response_2_EPs}
\end{figure}

Next, we evaluate the expression  $\braket{J_m^l|\left(\hat{H}_{EP}-\Omega_{EP}^{(1)}\right)|I_k^r}$ for $k,m>1$. By following the procedure described above and by using Eqs.\:(\ref{Eq:left1}) and (\ref{Eq:right2}), we find $\left(\Omega_{EP}^{(2)}-\Omega_{EP}^{(1)}\right)\braket{J_m^l|I_k^r}=\chi_m^{(1)}\braket{J_{m-1}^l|I_k^r}-\chi_k^{(2)}\braket{J_m^l|I_{k-1}^r}$. The above formula can be used recursively to replace $\braket{J_{m-1}^l|I_k^r}$ and $\braket{J_m^l|I_{k-1}^r}$ by generalized eigenvectors having lower order indices to arrive at $\braket{J_m^l|I^r_k}=\sum_{m'=2}^m c_{m'}^{(1)}\braket{J_{m'}^l|I_1^r}+\sum_{k'=1}^k c_{k'}^{(2)}\braket{J_1^l|I_{k'}^r}$. But the right hand side of this last expression is zero, and hence $\braket{J_m^l|I^r_k}=0$. Similarly,  $\braket{I_k^l|J_m^r}=0$. In other words, the generalized eigenvectors of the two different EPs form a biorthogonal set of vectors.\par

Equipped with this information, we can consider any arbitrary input signal $\sum_{n=1}^{N-M_1-M_2}c_n\ket{\psi_n^r}+\sum_{m=1}^{M_1}d_m\ket{J_m^r}+\sum_{k=1}^{M_2}f_k\ket{I_k^r}$, and repeat the same procedure as in supplementary note \ref{Appendix:response derivation}, to obtain:

\begin{align} \label{Eq:Greens_2EPs}
	\hat{G}_{EP}\ket{u}=&\left(\sum\limits_{n=1}^{N-M_1-M_2}\frac{\ket{\psi_n^r}\bra{\psi_n^l}}{\omega-\Omega_n}+\nonumber
	\sum\limits_{m=1}^{M_1}\sum\limits_{m'=m}^{M_1}\alpha_{m'}^{(m)}\frac{\ket{J_m^r}\bra{\tilde{J}_{m'}^l}}{\left(\omega-\Omega_{EP}^{(1)}\right)^{m'-m+1}}+\right. \nonumber\\
	&\qquad \left. \sum\limits_{k=1}^{M_2}\sum\limits_{k'=k}^{M_2}\beta_{k'}^{(k)}\frac{\ket{I_k^r}\bra{\tilde{I}_{k'}^l}}{\left(\omega-\Omega_{EP}^{(2)}\right)^{k'-k+1}} \right) \ket{u}
\end{align}

\noindent where $\alpha_{m}^{(m)}=1, \ \alpha_{m'}^{(m)}=\alpha_{m'-1}^{(m)}\chi_{m'}^{(1)}$ and  $\beta_{k}^{(k)}=1, \ \beta_{k'}^{(k)}=\beta_{k'-1}^{(k)}\chi_{k'}^{(2)}$. As discussed in supplementary note \ref{Appendix:bi-orthogonality} for the case of only one EP, the vectors $\bra{\tilde{J}_m^l}$ and $\bra{\tilde{I}_k^l}$ span the dual domains associated with $\mathcal{D}_{EP}^{(1,2)}$ and satisfy the relations  $\braket{\tilde{J}_m^l|J_n^r}=\delta_{n,m}$ and $\braket{\tilde{I}_m^l|I_n^r}=\delta_{n,m}$. The general structure described by the above expansion is shown schematically in supplementary Fig.\:\ref{Fig:Response_2_EPs}.

\subsection{Photonic simulations}\label{Appendix:Photonic simulation}
Here we explain the details of the simulations associated with Fig.\:6\textbf{a} in the main text. The outer radius and width of the microring resonator was chosen to be $R = 5\: \upmu$m and  $w = 0.25 \:\upmu$m, respectively. The material refractive index is assumed to be $n_r = 3.47$ and the background index was taken as $n_b = 1.44$ (relevant to silicon photonics). The straight waveguides are of the same material and width as the ring. The edge-to-edge separation between the ring and each of the  waveguides is $d = 0.2\: \upmu$m. By first using an add-drop configuration (in the absence of the mirror) and plotting the drop signal as a function of frequency (see supplementary Fig.\:\ref{Fig_gamma}), we evaluate the value of the decay rate: $\gamma=123\:$GHz. From the same figure, we also find that $\omega_o=1217 \times 10^{12} \: \text{sec}^{-1}$, corresponding to $f_o=193.7 \: \text{THZ}$ or equivalently $\lambda_o=1548\:$nm for the TE optical mode. The mirror is implemented by using  a 100-nm-thick silver layer. The absolute value of the amplitude reflection coefficient of the  mirror is obtained by simulating the waveguide-mirror system in the absence of the ring resonator, and is found to be $|r|\approx 0.98$. The stored energy enhancement $\eta$ is calculated by integrating the time averaged energy density over the ring resonator volume for resonant excitation from ports $\text{P}_2$ and $\text{P}_1$ under steady-state conditions (see Fig.\:6\textbf{b} in the main text). All simulations were performed by using 2D finite element method available from COMSOL software package.

\begin{figure}
	\includegraphics[scale=1]{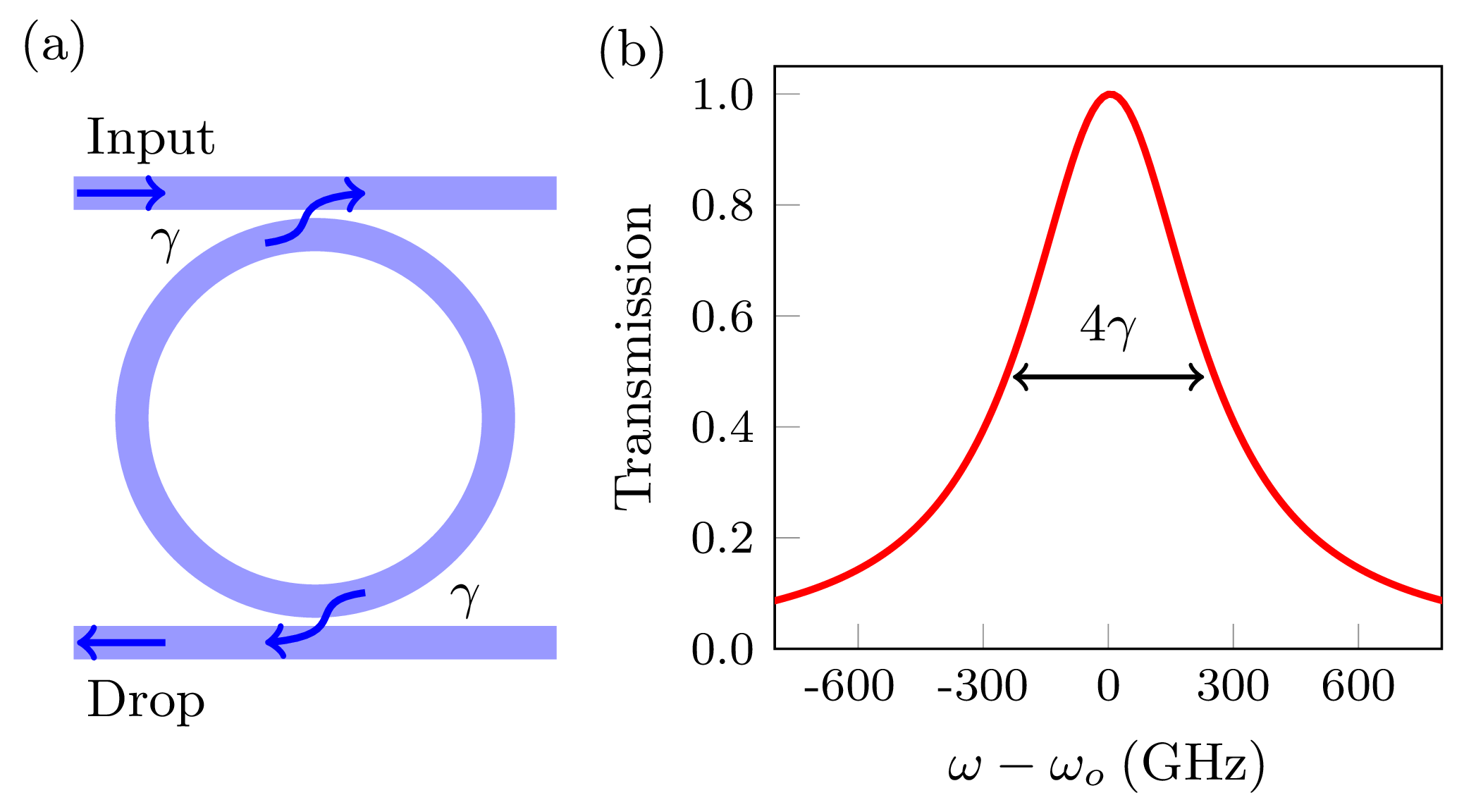}
	\caption{(a) Add-drop filter configuration for evaluating the resonant frequency and decay rate of a microring resonator. The design parameters of the ring resonator and the waveguides are listed in the text. (b) Optical power transmission as a function of frequency detuning from the resonant frequency (estimated from the simulation data to be  $f_o=193.7 \: \text{THZ}$) as obtained by frequency domain finite element numerical analysis. From the plot, we estimate that $\gamma=123\:$GHZ.}
	\label{Fig_gamma}
\end{figure}
\newpage

\bibliography{Reference}

\end{document}